# A Multiscale Framework for Defining Homeostasis in Distal Vascular Trees: Applications to the Pulmonary Circulation

Vasilina Filonova[1], Hamidreza Gharahi[2], Nitesh Nama[1], Seungik Baek[2], C. Alberto Figueroa[1,3]

**Abstract**
Coupling hemodynamics with vessel wall growth and remodeling (G&R) is crucial for understanding pathology at distal vasculature to study progression of incurable vascular diseases, such as pulmonary arterial hypertension. The present study is the *first* modeling attempt that focuses on defining homeostatic baseline values in distal pulmonary vascular bed via, a so-called, *homeostatic optimization*. To define the vascular homeostasis and total hemodynamics in the vascular tree, we consider two time-scales: a cardiac cycle and a longer period of vascular adaptations. An iterative homeostatic optimization is performed at the slow-time scale and incorporates: an extended Murray's law, wall metabolic cost function, stress equilibrium, and hemodynamics. The pulmonary arterial network of small vessels is represented by a fractal bifurcating tree. The pulsatile blood flow is described by a Womersley's deformable wall analytical solution. A vessel wall mechanical response is described by the constrained mixture theory for an orthotropic membrane and then linearized around mean pressure. Wall material parameters are characterized by using available porcine pulmonary artery experiments and human data from literature. Illustrative examples for symmetric and asymmetric fractal trees are presented to provide homeostatic values in normal subjects. We also outline the key ideas for the derivation of a temporal multiscale formalism to justify the proposed one-way coupled system of governing equations and identify the inherent assumptions. The developed framework demonstrates a potential for advanced parametric studies and future G&R and hemodynamics modeling in pulmonary arterial hypertension.

## 1. Introduction
### 1.1. Motivations

The pulmonary arterial hypertension (PAH) is a complex disorder defined by elevated mean pulmonary arterial pressure ($\geq$25 mmHg), which is associated with progressive wall thickening and stiffening of distal (small) pulmonary vessels. There are no unique symptoms associated with this disease, meaning a final diagnosis requires invasive and risky heart catheterization. This usually leads to delays in diagnosis and palliative treatments and, consequently, to high morbidity and mortality.

PAH pathology is manifested distally in smooth muscle hypertrophy, endothelial dysfunction, deposition of collagen, and increased elastolytic activities (Schermuly et al. 2011; Rabinovitch 2012). Such constituent level changes leads to wall thickening and stiffening, which can be described within a concept of growth and remodeling (G&R) of biological tissues (Ambrosi et al. 2011). A baseline for tissue G&R can be presented via a homeostasis process (Humphrey, Dufresne, and Schwartz 2014).

Pulmonary vascular hemodynamics and tissue changes have different length scales (tissue, cell, subcellular structure) and different time scales (cardiac cycle in seconds, synthesis and turnover of wall constitutions from seconds to months, and vascular growth adaptation from days to years), **Fig. 1**.

There are, however, difficulties in obtaining direct measurements of hemodynamics and tissue G&R of the distal pulmonary arteries *in-vivo*, which is essential for investigating PAH. For instance, standard medical imaging modalities do not provide information on the arterial tissue structure and hemodynamics.

[1] Section of Vascular Surgery, Department of Surgery, University of Michigan, Ann Arbor, MI 48109, USA
[2] Department of Mechanical Engineering, Michigan State University, East Lansing, MI 48824, USA
[3] Departments Biomedical Engineering, University of Michigan, Ann Arbor, MI 48109, USA

Therefore, this work aims to develop data-driven computational models for G&R and hemodynamics to enhance our understanding of the progression of PAH.

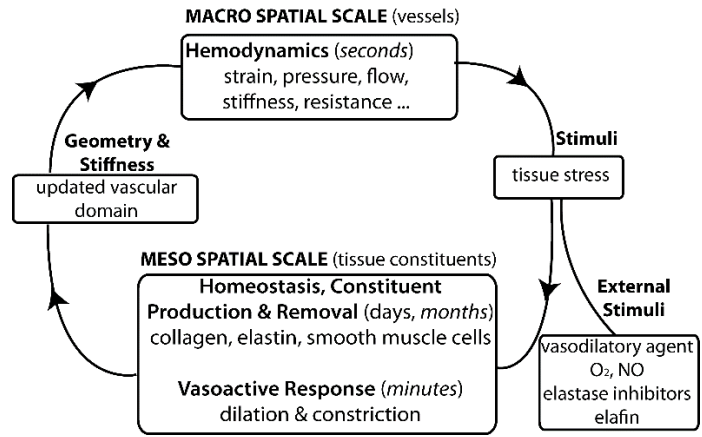

**Fig. 1** Information exchange for multiscale models that account for hemodynamics, tissue homeostasis (G&R) at macro- and meso-spatial scales as well as at slow- and fast-time scales.

### 1.2. Background

Key components of the computational model in a vascular tree must include: 1) fluid-solid-interaction (FSI), 2) G&R and homeostatic baseline, and 3) fluid-solid-growth (FSG).

#### *1.2.1. Fluid-Solid-Interaction in a Vascular Tree*

Examples of computational modeling in an arterial tree with FSI can be found in many works (Qureshi et al. 2014; Mittal et al. 2005; Huo and Kassab 2007; van de Vosse and Stergiopulos 2011; Olufsen et al. 2000). The vascular tree structure plays an important role for the circulation hemodynamics, because a network geometry defines downstream resistance. Hemodynamics (and FSI) in distal vasculature is usually described by 1D blood flow theory (Hughes and Lubliner 1973; van de Vosse and Stergiopulos 2011; Womersley 1955, 1957). A 1D model can be coupled to a 3D model of a proximal vasculature via boundary conditions presented by lumped parameter methods (Figueroa et al. 2006; Vignon-Clementel et al. 2006).

Building physiologically realistic and complete tree morphometry (geometry) of small vasculature is challenging due to data acquisition and modeling complexity (Burrowes et al. 2005). Depending on data sources (images, tree casts) and the desired accuracy, a pulmonary vascular network can be reconstructed in low- and high-fidelity models. Numerous models include: fractal trees (Olufsen et al. 2000; Qureshi et al. 2014; Ionescu et al. 2009; Schreiner et al. 2006); Strahler ordering technique (Huang et al. 1996; Jiang, Kassab, and Fung 1994; Tamaddon et al. 2016); volume-filling branching algorithm (Burrowes et al. 2005); etc. The fractal tree is one of the simplest modeling approaches which makes it appealing for mathematical modeling. Noteworthy, the lung airway structure, which pulmonary arteries follow distally, is fractal as was inferred by Mandelbrot (Mandelbrot 1983) – the developer of the fractal theory. A fractal tree model was adopted in pulmonary arteries in (Krenz, Linehan, and Dawson 1992; Bennett et al. 1996). In core papers of Olufsen and coworkers (Olufsen et al. 2012; Qureshi et al. 2017; Olufsen et al. 2000; Qureshi et al. 2014), a radii-split fractal rule was proposed to represent various vascular networks and model 1D blood flow in systemic and pulmonary circulations.

#### *1.2.2. Growth and Remodeling and Homeostatic Baseline in a Vascular Network*

A biomechanical description of tissue G&R at a constitutive level is developed in works of Humphrey and co-workers (Humphrey and Rajagopal 2002; Baek, Valentín, and Humphrey 2007). This approach uses the constrained mixture theory (Humphrey and Rajagopal 2002) to model mechanical behavior of an arterial wall. This theory allows one to consider the evolution of individual wall constituents (elastin, collagen, smooth muscle cells) by incorporating biologically driven mass production and removal as a function of biomechanical stimuli such as changes in mechanical stresses (wall shear stress and hoop stress). The G&R framework postulates that in a healthy condition, the blood vessel wall is maintained in a mechanical homeostatic state determined by cellular mechanotransduction. The homeostatic stresses, however, are not constant throughout an arterial network as shown in (Pries, Reglin, and Secomb 2005). While the G&R framework does not explain the variability of homeostatic values throughout the arterial network, energy-based approaches, such as Murray's law, can provide a such an insight.

Murray (Murray 1926) suggested that *in-vivo* design of the arterial network is determined by a balance between the cost of maintaining blood volume and power needed to overcome viscous forces. While Murray's law provided a model to the arterial design (radii and angle split at bifurcations), one of its limitations is that it does not consider the metabolic power spent by vascular wall, endothelial, smooth muscle, and fibroblast cells. Later, several generalizations of Murray's law appeared in (Uylings 1977; Kassab and Fung 1995; Zhou, Kassab, and Molloi 1999; Taber 1998; Y. Liu and Kassab 2007). These studies aimed to find important morphometric and hemodynamic relationships in blood vessels. Specifically, Taber (Taber 1998) proposed a cost function that included arterial wall basal metabolism and active tension. In his study, an extension of Murray's law was used to assess a hypothesis in which the homeostatic wall shear stress is regulated by the blood pressure. Alternatively, Liu and Kassab (Y. Liu and Kassab 2007) incorporated the same idea into a coronary arterial tree (Zhou, Kassab, and Molloi 1999) and concluded that total metabolic consumption is proportional to the entire arterial tree volume. More recently, Lindström et al. (Lindström, Satha, and Klarbring 2015; Satha, Lindström, and Klarbring 2014) extended Murray's law by including nonlinear mechanics of composite artery walls and introducing a goal-function-based G&R.

Although the concept of stress-mediated G&R has been successful in explaining and predicting vascular adaptation in physiological and pathological conditions, previous studies (Baek, Rajagopal, and Humphrey 2006; Baek, Valentín, and Humphrey 2007; Farsad et al. 2015) were mostly limited to large systemic arteries. This framework, however, can be extended to study multiscale phenomena of PAH addressing the need to model G&R process in an arterial network. As the first step to such model, it is crucial to determine the homeostatic state for pulmonary vasculature.

### 1.2.3 *Fluid-Solid-Growth*

The concept of combining the computational G&R model with hemodynamics (FSI) for an individual vessel was shown in the FSG framework (Figueroa et al. 2009). The FSG framework proposed an iterative coupling between a slow-time G&R model of a basilar artery with fast-time hemodynamics. In summary, by assuming small deformations of the vessel wall over a cardiac cycle, the material stiffness of the wall was linearized at mean pressure in the intermediate configuration, using the theory of small on large (Baek et al. 2007). Alternatively, the hemodynamic forces (mean wall shear stresses and pressure) from FSI modeling were used as biomechanical stimuli for the G&R modeling.

## 1.3. Outline

In the present study, we propose a framework for modeling the homeostatic baseline and hemodynamics of the distal pulmonary vasculature. Inspired by the fractal nature of pulmonary arteries, we initialize a fractal tree based on the radii-split rule (Olufsen et al. 2000). Then, we develop a numerical approach to define the homeostatic baseline of the arterial tree that will update the fractal rule based on the extended Murray's law (Lindström, Satha, and Klarbring 2015) and slow-time hemodynamics. Finally, we analytically calculate the pulsatile hemodynamics in the arterial tree using deformable wall Womersley's theory (Womersley 1957) adapted to the orthotropic material and tree bifurcations (Nichols, O'Rourke, and Vlachopoulos 2011; van de Vosse and Stergiopulos 2011).

The manuscript is organized as follows. We start with the mathematical formulation of the two-scale FSI problem in a tree (**Section 2**). In **Section 2.1**, we introduce governing equations for one vessel. In **Section 2.2**, the Womersley's deformable wall theory and Laplace law are incorporated to the system of governing equations. In **Section 2.3**, the hemodynamic analytical solution for a pulmonary arterial tree is described. Then in **Section 3**, we introduce the homeostatic optimization in three parts: initialization of a fractal tree structure (**Section 3.1**), description of the minimization problem in one vessel (**Section 3.2**), and an iterative procedure for homeostatic optimization in a pulmonary tree (**Section 3.3**). Finally, in **Section 4**, we demonstrate the numerical examples describing the parameterization (**Section 4.1**) followed by results for a symmetric (**Section 4.2**) and asymmetric (**Section 4.3**) arterial tree. We close the manuscript with the **Discussion** and **Appendix** sections.

## 2. Fluid-Solid Interaction

In this section, we describe the FSI formulation for an arterial tree. We distinguish two time scales (fast and slow): a homeostatic analysis operates at the slow-time scale, while oscillations of hemodynamic and wall deformation variables are estimated over the cardiac cycle at the fast-time scale. We propose the governing equations for fluid (blood) and solid (deformable wall) in the two time scales. In **Appendix A**, we introduce key ideas of the time-scale separation formalism and suggest a scenario for justifying the one-way coupling system of governing equations that we utilize in the present study. The proposed formulation is analogous with the FSG framework (Figueroa et al. 2009), though it is presented with a case of linear blood flow. The vessel wall is considered a linearly elastic membrane.

### 2.1. Two-scale Governing Equations

The slow-time system describes fluid and solid equations to find the slow-time fluid velocity $v^s$, pressure $p^s$, and wall displacements $u^s$. Fluid equations include the divergence-free velocity field and static equilibrium between the pressure gradient and viscous forces. The coordinate system $x$ is related to the intermediate configuration (**Appendix A**, **Fig. 16**). The solid equation describes an equilibrium for the slow-time wall stress and a body force:

$$\text{slow - time: } \nabla_x \cdot v^s = 0, \qquad \nabla_x p^s = \mu \nabla_x^2 v^s,$$
$$0 = \nabla_x \cdot P^s_{solid}(u^s) + \mathbf{B}^s, \qquad (1)$$

where $\mu$ is the constant viscosity of blood, $P^s_{solid}$ is the averaged first Piola-Kirchhoff stress tensor, and $\mathbf{B}^s$ is an averaged body force vector. The wall stress tensor is expressed via the strain energy function and a constrained mixture model (**Appendix D**). The interaction between the fluid and vessel wall can be postulated using Laplace law.

In addition, the fast-time system describes fluid and solid equations to find the fast-time fluid velocity $v^f$, pressure $p^f$, and displacements $u^f$, which oscillate around a given slow-time state. The equations include the divergence-free velocity, fluid balance of momentum with linear inertia force, and elastodynamics equation for the solid:

$$\text{fast - time: } \nabla_x \cdot v^f = 0, \qquad \rho_{fluid} \left. \frac{\partial v^f}{\partial t} \right|_x = -\nabla_x p^f + \mu \nabla_x^2 v^f,$$
$$\rho_{solid} \left. \frac{\partial^2 u^f}{\partial t^2} \right|_x = \nabla_x \cdot \left. P^f_{solid} \right|_s + \mathbf{B}^f, \qquad (2)$$

where $\rho_{fluid}$ and $\rho_{solid}$ are the blood and wall density, respectively. $\left. P^f_{solid} \right|_s$ is the first Piola-Kirchhoff stress, which is defined as a linearization of the nonlinear constrained mixture model around the slow-time state. $\mathbf{B}^f$ is the fast-time body force vector. The fluid-solid interaction can be presented via a fictitious body force in the membrane wall (Womersley 1955; Figueroa et al. 2006). Noteworthy, the equations (1) and (2) are one-way coupled in time (from slow-to-fast) as the result of the proposed kinematics in **Appendix A**. This one-way coupling allows us to obtain a computationally feasible system of governing equations that can be solved efficiently for the entire arterial tree.

## 2.2. Womersley Theory and Laplace Law in a Vessel

The flow solution to the slow-time (steady-state) system of equations (1) in an artery can be presented by Poiseuille law:

$$q^s = \frac{\pi R^4 \Delta p^s}{8 \mu L}, \tag{3}$$

where $q^s$ is the steady-state flow rate, $\Delta p^s$ is the steady-state pressure drop along the vessel, $R$ is the vessel lumen radius, and $L$ is the vessel length. The equilibrium equation for the vessel wall in (1) can be reduced to the Laplace law for the cylindrical membrane, subjected to the distension pressure $p^s$ (Humphrey 2002):

$$\left. T_{\theta\theta} \right|_s = p^s R . \tag{4}$$

where $\left. T_{\theta\theta} \right|_s$ is the membrane Cauchy stress in the circumferential direction defined at the slow-time state, **Appendix E**.

Furthermore, the solution to the fast-time system of equations (2) in an artery can be described by Womersley's theory of the oscillatory flow in deformable tubes (Womersley 1957). This theory provides an analytical solution for the fast-time periodic variables at a given frequency $\omega$:

$$p^f = P e^{i\omega(t-z/c)}, \quad q^f = Q e^{i\omega(t-z/c)}, \quad Q = \frac{\pi R^2 P}{c \rho_{fluid}} (1-g),$$

$$v_z^f = \frac{P}{c \rho_{fluid}} \left( 1 - \frac{J_0(\Lambda r / R)}{\Lambda J_0(\Lambda)} \right) e^{i\omega(t-z/c)}, \tag{5}$$

where $q^f$ is the fast-time flow rate, $v_z^f$ is the fast-time longitudinal velocity, and $P$ and $Q$ are fast-time pressure and flow in the frequency domain, $z$ is a coordinate along the vessel. $J_0$ and $J_1$ are the Bessel functions of the first kind; $g = 2J_1(\Lambda)/(\Lambda J_0(\Lambda))$ and $\Lambda = i^{3/2} \alpha^{Wom}$ with the Womersley number defined as $\alpha^{Wom} = R\sqrt{\omega \rho_{fluid} / \mu}$.

Moreover, the pulse wave velocity $c$ is given by the Womersley's solution (**Remark 1**). In the current study, however, we consider the orthotropic membrane stiffness, and therefore, accordingly modify the pulse wave velocity:

$$c = \sqrt{\frac{(1-g) h \left. \mathcal{A}_{\theta\theta} \right|_s}{2 R \rho_{fluid}}}. \tag{6}$$

where $\left. \mathcal{A}_{\theta\theta} \right|_s$ is the main component of the stiffness tensor for a longitudinally constrained vessel, **Appendix A**.

**Remark 1**. Pulse wave velocity is a clinically important cardiovascular metric to quantify the structural stiffness of the vessel wall *in-vivo*. In the case of an isotropic elastic material, the pulse wave velocity given in (Womersley 1957) is $c = c^{MK}\sqrt{(1-g)/(1-\sigma^2)}$, where $c^{MK} = \sqrt{hE/(2R\rho_{fluid})}$ is the Moens-Korteweg wave speed for an inviscid fluid with Poisson ratio $\sigma$ and Young's modulus $E$. Bergel (Bergel 1961) proposed the wave speed for the elastic arterial wall as $c = c^{MK}/\sqrt{(1-\sigma^2)}$, consistent with Womersley's solution at $g = 0$ (inviscid flow). For an elastic, anisotropic, longitudinally constrained shell, Mirsky (Mirsky 1978) derived the pulse wave velocity as $c = \sqrt{h\mathcal{A}_{\theta\theta}/(2R\rho_{fluid})}$, which reduces to the wave speed shown in (Bergel 1961) for the isotropic case. Noteworthy, the wave speed in (6) is similar to (Mirsky 1978) when $g = 0$.

The total pressure $p$ and flow rate $q$ in the time domain are obtained by applying Fourier series to the solutions (5) for multiple frequencies $\omega_n = 2\pi n/T$ superimposed by the slow-time solution (3), (4) at zero-frequency:

$$p = p^s + \sum_{n=1}^{\infty} P_n e^{i\omega_n(t-z/c_n)}, \quad q = q^s + \sum_{n=1}^{\infty} Q_n e^{i\omega_n(t-z/c_n)}, \tag{7}$$

where $P_n$ and $Q_n$ are frequency dependent coefficients of oscillatory fields.

Moreover, we compute the characteristic impedance $Z^c(\omega)$ and hydraulic resistance:

$$Z^c(\omega) = \frac{P}{Q} = \frac{c\rho_{fluid}}{\pi R^2 (1-g)} = \frac{1}{\pi R^2}\sqrt{\frac{h\rho_{fluid}\mathcal{A}_{\theta\theta}\big|_{p^s}}{2R(1-g)}}, \quad res = \frac{\Delta p^s}{q^s} = \frac{8\mu L}{\pi R^4}. \tag{8}$$

The characteristic impedance in the time domain is $z^c = p^s(0)/q^s + res + \sum_{n=1}^{\infty} Z_n^c e^{iw_n t}$. Similarly, we can compute the input impedance $Z^{inp}(\omega) = P/Q\big|_{z=0}$ and terminal impedance $Z^T(\omega) = P/Q\big|_{z=L}$.

### 2.3. Hemodynamics in a Tree

In this section, we extend the hemodynamics solution for one vessel to an entire arterial tree. We consider the pulmonary arterial tree as a 1D bifurcating network, where each vessel can be characterized by a radius and length. Similarly to the previous section, the total hemodynamics is the summation of the slow-time and fast-time hemodynamics. The slow-time hemodynamics in the arterial tree is computed using Poiseuille flow for each vessel flow conservation and pressure continuity for each bifurcation (**Appendix B**), and boundary conditions at the tree inlet and outlets. Given the arterial tree and vessel wall stiffness, we can compute the fast-time hemodynamics using the analytical solutions (7).

Each bifurcation is a reflection site for a traveling wave. Due to the system linearity, the fast-time pressure and flow can be additively decomposed to forward and backward waves:

$$P = P_{forw} + P_{back}, \quad P_{forw} = H_{forw} e^{-i\omega z/c}, \quad P_{back} = H_{back} e^{i\omega z/c}$$
$$Q = Q_{forw} + Q_{back}, \quad Q_{forw} = \frac{H_{forw}}{Z^c} e^{-i\omega z/c}, \quad Q_{back} = -\frac{H_{back}}{Z^c} e^{i\omega z/c} \tag{9}$$

with constant coefficients $H_{forw}$ and $H_{back}$ for individual harmonics. The characteristic impedance (8) and pulse wave velocity (6) are the same for both forward and backward waves (van de Vosse and Stergiopulos 2011): $Z^c = P_f/Q_f = -P_f/Q_f$. Therefore, we compute the reflection coefficient at the end of an individual vessel (Taylor 1957; Avolio 1980):

$$\Gamma = \frac{Z^T - Z^c}{Z^T + Z^c} = \frac{P_{back}}{P_{forw}}\bigg|_{z=L}. \tag{10}$$

The reflection coefficient varies between -1 and 1. Particularly, $\Gamma = 0$ indicates no wave reflection in the vessel, i.e. matching impedance $Z^T = Z^c$. The open-end type (or negative) wave reflection is associated with $\Gamma = -1$, i.e., $Z^T = 0$ or $P\big|_{z=L} = 0$. In addition, the closed-end type (or positive) wave reflection is associated with $\Gamma = 1$ for $Z^T = \infty$ or $Q\big|_{z=L} = 0$. See **Remark 2** for the type of reflection found in pulmonary circulation.

Using $\Gamma$ and $Z^c$, the impedance can be obtained as a function along the vessel:

$$Z(z,\omega) = \frac{P}{Q}(z,\omega) = Z^c \frac{1 + \Gamma e^{-i\omega 2(z-L)/c}}{1 - \Gamma e^{-i\omega 2(z-L)/c}}. \tag{11}$$

Also, we can compute the oscillatory pressure and flow along the vessel using (9) and (10):

$$P(z,\omega) = H_{forw} e^{-i\omega z/c}(1 + \Gamma e^{i\omega 2(z-L)/c}), \quad Q(z,\omega) = \frac{H_{forw}}{Z^c} e^{-i\omega z/c}(1 - \Gamma e^{i\omega 2(z-L)/c}). \tag{12}$$

Each bifurcation is characterized with a parent vessel and two daughter vessels. Considering the conservation of flow and pressure continuity at each bifurcation, we can obtain the terminal impedance in the parent vessel:

$$Z_p^T = \left(\frac{1}{Z_{d1}^{inp}} + \frac{1}{Z_{d2}^{inp}}\right)^{-1}, \tag{13}$$

where the input impedance for each daughter vessel is computed from (11):

$$Z^{inp} = Z^c \frac{1 + \Gamma e^{-i\omega 2L/c}}{1 - \Gamma e^{-i\omega 2L/c}}. \tag{14}$$

Given the flow at the root vessel and the terminal pressure and reflection coefficient at the outlets, we use the bifurcation relation (13) to compute the input impedance (14) and reflection coefficient (10) from bottom-to-top. Finally, we can reconstruct the pressure and flow at each vessel of the tree from top-to-bottom, using (12). The algorithmic details are in **Appendix B and C**.

**Remark 2.** Using wave-intensity analysis, (Hollander et al. 2001) demonstrated that the normal pulmonary arterial circulation (in dogs) is characterized by negative wave reflections. The primary factor for creating the open-end reflection type in the pulmonary arterial system is likely the large increase in the total arterial cross-sectional area over a short distance. The magnitude of the negative reflection increases when the daughter-to-parent area ratio $a_d / a_p > 1.2$. Specifically, in the pulmonary circulation, the area ratio is $1.2 < a_d / a_p < 1.3$.

## 3. Homeostatic Optimization

In this section, we describe a method to estimate the homeostatic baseline characteristics of an entire arterial tree based on an extension of Murray's and the constrained mixture models of the arteries. In particular, the proposed optimization is an energy-based minimization problem that operates at the slow-time scale. The optimization constraints include the slow-time hemodynamics solution and Laplace law, (3) and (4). This procedure will determine the homeostatic radii distribution over the pulmonary arterial tree as well as the steady-state pressure and flow.

In the next sections, first, we present the initialization of the arterial tree structure. Second, we proceed to describe the optimization problem for a single vessel. Finally, we expand the optimization problem to the arterial tree.

## 3.1. Tree Initialization

In the current study, we proceed with a fractal tree representation of the pulmonary arterial network. In general, the homeostatic optimization can be applied to any reconstruction method for the pulmonary arterial tree, including Strahler ordering technique (Huang et al. 1996).

An initial bifurcating fractal tree is constructed using the radii-split rule given in (Olufsen et al. 2000): $R_{d1} = \alpha R_p$, $R_{d2} = \beta R_p$ for the parent vessel radius $R_p$ and daughter vessels radii $R_{d1}$ and $R_{d2}$. Given the exponent $\xi$ for the Murray's law $R_p^\xi = R_{d1}^\xi + R_{d2}^\xi$ and the area ratio $a_d/a_p = (R_{d1}^2 + R_{d2}^2)/R_p^2$, the asymmetry ratio $\gamma = R_{d2}^2/R_{d1}^2$ can be found by solving the equation $1 + \gamma = (1 + \gamma^{\xi/2})^{2/\xi} a_d/a_p$. This leads to the following scaling parameters: $\alpha = (1 + \gamma^{\xi/2})^{-1/\xi}$ and $\beta = \alpha\sqrt{\gamma}$. Thus, for each generation $k$ the vessel radius is $R(i,j) = \alpha^{i-1}\beta^{j-1}R_1$, where $R_1$ is the radius of the tree root vessel and $i + j - 1 = k$, (**Fig. 2**). The vessel length is defined as a function of radius $L(R)$.

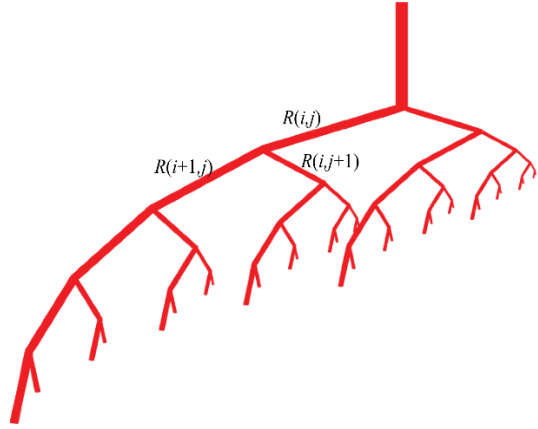

**Fig. 2** An illustrative example of initial geometrical, fractal tree with 6 generations and the radius rule $R(i,j) = \alpha^{i-1}\beta^{j-1}R_1$ obtained with $\xi = 2.76$ and $\gamma = 1.16$. The bifurcation angles and vessel lengths are adjusted for visualization.

To preserve the asymmetry during energy optimization, the asymmetric tree has to be initialized with pruning that yields branch termination at different generations. To enforce this, the bifurcations are pruned when the radius reaches the minimum value $R_{min}$.

## 3.2. Cost Function in a Vessel

Following the study by (Lindström et al. 2015), we assume that the blood vessel wall composition and geometry strive to optimize the energy consumption. This energy includes the metabolic demand for maintaining the volume of the blood in a vessel and the power needed to overcome hydraulic resistance. In addition, we have to account for the metabolic demand for the maintenance (production and removal) of vessel wall constituents. We assume that the homeostatic state is governed by such an optimization rule that can be defined for each individual vessel.

First, the metabolic power needed for blood supply is proportional to the blood volume that needs to be sustained, with a factor $\vartheta^{blood}$. This metabolic power per unit length can be written as $C_{blood} = \vartheta^{blood}\pi R^2$. Second, the power per unit length needed to overcome the resistance of Poiseuille flow (viscous drag forces) is $C_{drag} = 8\mu q^{s2}/(\rho_{fluid} \pi R^4)$. Third, the metabolic cost per unit length of the arterial wall constituents is assumed to be proportional to the mass of each constituent: $C_{wall} = (2\pi R/\rho_{solid}) \sum_i \vartheta^i M_R^i$. Here $M_R^i$ is the mass per unit reference area of each constituents for $i \in \{el, smc, col\}$ representing elastin, smooth muscle cells (SMCs), and collagen fibers, respectively. The constant $\vartheta^i$ is the metabolic energy cost per unit volume and $R$ is the vessel radius in homeostatic conditions. The metabolic cost of smooth muscle cells (SMC) $\vartheta^{smc}$ includes both the metabolic cost of maintenance $\vartheta_{maint}^{smc}$ and active tension $\vartheta_{act}^{smc}$.

Finally, the total energy cost per unit length for an individual blood vessel can be summarized as:

$$C(M_R^i, R; q^s) = C_{blood} + C_{drag} + C_{wall} = \vartheta^{blood} \pi R^2 + \frac{8\mu q^{s2}}{\rho_{fluid} R^4} + \frac{2\pi R}{\rho_{solid}} \sum_i \vartheta^i M_R^i. \tag{15}$$

The energy cost function in (15) has to be minimized with Laplace law as a constraint. To incorporate Laplace law in the cost function, we rewrite it in terms of the mass of vessel wall constituents. Accordingly, we define the total mass as $M_R^{total}$ and mass fractions of elastin, SMCs, and collagen fibers as $v^{el} = M_R^{el}/M_R^{total}$, $v^{smc} = M_R^{smc}/M_R^{total}$, and $v^{col} = \sum_k v^k$, respectively, with $k$-th collagen-fiber-family mass fraction $v^k = M_R^k/M_R^{total}$. Then, the circumferential membrane stress can be written as (**Appendix D**):

$$T_{\theta\theta} = \frac{M_R^{total}}{(1-\phi_f)\rho_{solid}\lambda_\theta \lambda_z} \sum_i v^i \sigma_\theta^i, \tag{16}$$

where the parameter $\phi_f$ is the volume fraction of the interstitial fluid, usually taken as 70%. Furthermore, $\sigma_\theta^i$ is the circumferentially acting part of the Cauchy stress tensor $\boldsymbol{\sigma}^i$, defined by:

$$\boldsymbol{\sigma}^i = \frac{1}{h^i}\mathbf{T}^i, \quad h^i = \frac{M_R^i}{(1-\phi_f)\rho_{solid}\lambda_\theta \lambda_z}, \tag{17}$$

where $h^i$ is the effective thickness of each constituent in the vessel wall. Noteworthy, the total thickness of the arteries $h$ is defined as the sum of the constituent-wise thicknesses. Variations in hemodynamics under physiological conditions change the arterial wall stresses which leads to adaptive responses from the vascular wall cells (endothelial cells, SMCs, and fibroblasts). These responses include changes in SMC tone and/or extracellular matrix turnover. In the equilibrium, i.e. homeostatic state, the mechanical stress $\sigma_\theta^i$, sensed by vascular wall cells, remains constant. Although $\sigma_\theta^i$ is not directly measurable, its value can be computed from mechanical properties of the wall constituents (**Appendix D**).

Substituting (16) to Laplace law (4), we obtain the constraint:

$$p^s R = \frac{M_R^{total}}{(1-\phi)\rho_{solid}} \sum_i v^i \sigma_\theta^i. \tag{18}$$

Finally, embedding the constraint into the metabolic cost function for a single vessel (15), we can rewrite the optimization problem in terms of $R$:

$$C(R; \bar{p}^s, q^s) = \vartheta^{blood} \pi R^2 + \frac{8\mu q^{s2}}{\rho_{fluid} \pi R^4} + 2\pi(1-\phi)\bar{p}^s R^2 \frac{\sum_i \vartheta^i \nu^i}{\sum_i \nu^i \sigma_\theta^i}, \quad (19)$$

where slow-time hemodynamics, average pressure in a segment $\bar{p}^s$, and flow $q^s$, are iteratively computed for an arterial tree. This coupling is illustrated in the next section where the homeostatic optimization in a vascular tree is presented.

### 3.3. Iterative Process of Optimization

To solve the homeostatic optimization in an arterial tree, a global hemodynamic constraint has to be considered to determine $\bar{p}^s$ and $q^s$ for each vessel (**Appendix B**). This hemodynamic constraint is coupled with the optimization problem in an iterative manner (**Fig. 3** and **Appendix E**). First, the arterial tree is initialized as described in **Section 3.1**. Second, using this tree structure, the slow-time hemodynamics problem is solved for the entire tree. The boundary conditions for the hemodynamic problem are either pressure or flow, imposed at the inlet and outlets of the tree. Third, to update the vessel radii, the cost function minimization problem is solved in each segment using Newton-Raphson method. The second step, applied to the updated tree structure, and the third step are repeated until convergence is achieved (pressure remains constant). The algorithmic details of this implementation are given in **Appendix B, E**.

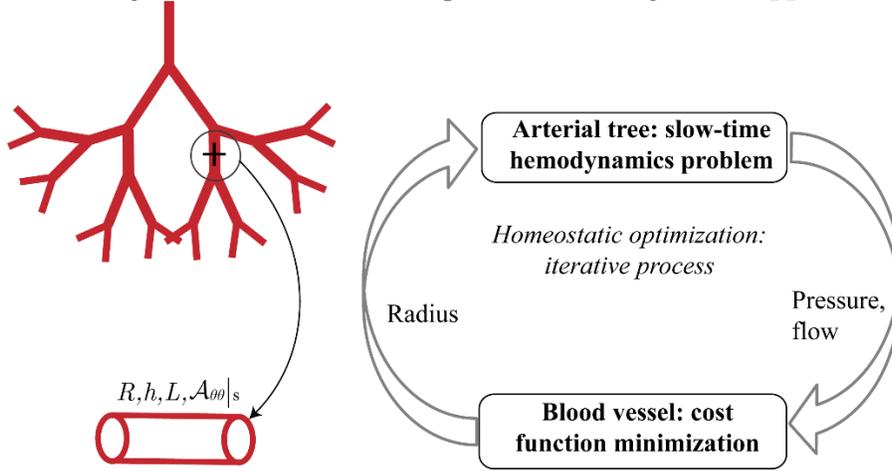

**Fig. 3** Schematic representation of the iterative process for homeostatic optimization: vessel stiffness, length, radius, wall composition and thickness are defined for individual vessels and the metabolic cost function is used to iteratively optimize for the radii using slow-time hemodynamics in the entire tree.

**Remark 3.** The fast-time hemodynamics in a tree (**Sections 2.2, 2.3**) is solved separately from the slow-time problem in a post-processing operation after the homeostatic optimization. This one-way coupling is facilitated through calculation of $\mathcal{A}_{\theta\theta}\big|_s$, after the homeostatic optimization provides the vessel wall mechanics. In accordance with this coupling, we present the homeostatic optimization followed by the results for the pulsatile hemodynamics.

## 4. Numerical Results

In this section, first, we present the results of applying our multiscale framework to an idealized symmetric pulmonary arterial tree. To demonstrate utility of the model, we consider two cases of constant and variable mass fractions along the arterial tree. Second, we illustrate the results of the framework application in an asymmetric arterial tree. For the asymmetric example, a full arterial tree is generated, then the distal vessels are pruned if the vessel radius is smaller than a specified minimum radius.

### 4.1. Parameters

The parameters of the models are listed in **Table 1**. The objective of the current examples is to estimate the homeostatic characteristics and hemodynamics of the pulmonary tree in the intermediate to small region of the vasculature (with radius of 100mm-5mm). For pulsatile hemodynamics, we prescribe an input flow waveform taken from human data in the main pulmonary artery (Zambrano et al. 2018) scaled to fourth generation downstream. In addition, the first nine harmonics are used to represent pressure and flow waveforms in the frequency domain. Furthermore, a terminal reflection coefficient is prescribed at the terminal vessels for fast-time hemodynamics. For the homeostatic optimization of the symmetric tree, we prescribe mean flow at the inlet and a terminal pressure (assumed to be close to the capillary pressure) at the outlets. In the asymmetric tree, however, we prescribe the inlet mean pressure (12 mmHg) and uniform flow at the outlets (mean flow/number of outlets). In all examples, $L(R)$ can be described using the relation given in (Olufsen et al. 2012) which approximates the human morphometric data from a single pulmonary arterial tree cast (Huang et al. 1996). We modified $L(R)$, scaling it by a factor of half, to obtain lengths closer to observations in the proximal vessels, (**Fig. 4b**).

Elastin in human arteries is mostly produced in early ages, and due to a large half-life, its content remains relatively constant over time in normal conditions. Therefore, in this study, the metabolic energy cost of elastin is neglected, $\vartheta^{el} = 0$. Moreover, the metabolic energy cost coefficients $\vartheta^i$ of maintenance of SMCs and collagen in the vessel wall are estimated to be of order 100-1000 W/m³ (Y. Liu and Kassab 2007). Although the metabolic cost of these constituents may vary throughout the arterial tree, we chose the value 1500 W/m³ for the entire tree. The metabolic cost of active tension in SMCs was reported in (Paul 1980) via the flux of adenosine triphosphates. Based on this study, Taber (Taber 1998) determined the value of active tension metabolic cost to be 0.00872 1/sec and proportional to the active tension. Since the active tension is proportional to the content of SMCs in the vessel wall, the metabolic cost can be assumed to be proportional to SMC content (**Appendix D**). Additionally, the metabolic energy requirements for sustaining blood in arteries are adopted from (D. Liu et al. 2012) based on the number of red blood cells, white blood cells, and platelets in a unit volume of blood and their corresponding oxygen consumption rate for a normal adult human subject.

The mechanical properties of the proximal vessels are calibrated against experimental data from inflation tests on right or left pulmonary arteries of pigs (Wang et al., n.d.) (**Table 1**). In the current study, we assume that the individual constituents preserve their mechanical behavior in different locations along the tree, though the deposition stretches are different. Therefore, intrinsic parameters of the constitutive laws for each individual wall constituents are considered constant (i.e., constant passive ($c_1 - c_5$) and active properties), whereas the pre-stretches ($G_\theta^{el}$, $G_z^{el}$, $G_h^{col}$, and $G_h^{smc}$) are calibrated in each vessel by matching the experimental thickness-to-diameter ratio (~7% to 16% from large arteries to microvessels) reported in (Li et al. 2012; Rol et al. 2017).

In Case 1, constant mass fractions for the constituents are prescribed in the entire arterial tree. The adventitial layer of the arterial wall was assumed to be comprised of 95% of collagen and 5% of elastin. Furthermore, Mackay and colleagues (Mackay et al. 1978) reported the mass fractions of the constituents in the medial layer of pulmonary arteries in healthy humans (**Table 1**). Finally, the mass fractions are estimated using the relative layer-wise thicknesses reported in (Chazova et al. 1995). In reality, however, the adult pulmonary arterial wall composition varies throughout the arterial tree (Elliott and Reid 1965). The variable composition of the vessel wall is reflected in Case 2 by changing the content of elastin and SMCs in the medial layer using data from (Chazova et al. 1995). Particularly, most of the arteries larger than 0.32 cm in diameter are elastic arteries endowed with multiple layers of elastic lamina. As arteries become smaller, their structure transitions from the elastic to muscular type over a range of 0.32-0.2 cm where the elastic layers fragment and are replaced by SMC (Elliott and Reid 1965). Arteries smaller than 0.2 cm have a muscular media with two distinctive internal and external elastic laminas (Hislop and Reid 1978). Although arteries smaller than 0.01cm are not included in this study, we note that the number of muscular arteries significantly drops when an arterial size approaches 0.01 cm and below. In this region, the vessels become partially- or non-muscular.

**Table 1** Model parameters for homeostatic optimization and pulsatile hemodynamics

| Parameter description | | Reference |
|---|---|---|
| *Initialization of the tree geometry* | | |
| Root vessel radius | 0.55 cm (symm); 0.36 cm (asymm) | (Olufsen et al. 2012) |
| Root vessel wall thickness | 7% of 0.55cm | |
| Terminal radius | 50 mm (symm); 180 mm (asymm) | (Olufsen et al. 2012) |
| Radial exponent $\xi$ | 2.72 | |
| Daughter-to-parent area ratio | 1.0 (symm); 1.2 (asymm) | |
| *Extended Murray's law* | | |
| Metabolic cost of collagen and SMCs ($\vartheta^{col}, \vartheta^{smc}_{maint}$) | 1500 W/m$^3$ | Based on (Y. Liu and Kassab 2007) |
| Metabolic cost of blood supply in human ($\vartheta^{blood}$) | 51.7 W/m$^3$ | (D. Liu et al. 2012) |
| Metabolic cost of active tension ($\vartheta^{smc}_{act}$) | 0.00872 sec$^{-1}$ | (Paul 1980) |
| *Vessel wall mechanical properties* | | |
| Wall density | 1060 kg/m$^3$ | |
| Case 1: Constant mass fractions (human data) | 77% collagen, 12% SMCs, 11% elastin | (Mackay et al. 1978) (Chazova et al. 1995) |
| Case 2: Variable mass fractions (elastic-muscular arteries) | 77-59% collagen, 12-39% SMCs, 11-2% elastin | (Elliott and Reid 1965) |
| Passive and active wall elasticity parameters | $c_1 = 28.83, c_2 = 178.59, c_4 = 24.51 N.m/kg$ $c_3 = 1.05, c_5 = 0.75$ $\alpha^c = 0, \pm 45, 90$ $\lambda_M = 1.2, \lambda_0 = 0.7$ $S = 20 kPa$ | (Wang et al., n.d.) |
| *Hemodynamics* | | |
| Input flow waveform, scaled by 1/8, (human) | mean flow 11.65 ml/s | (Zambrano et al. 2018) |
| Length-to-radius relation | $L = 6.2 R^{1.1}$ mm | |
| Symmetric tree: mean terminal pressure | 10 mmHg | (Qureshi et al. 2014) |
| Terminal reflection coefficient | -1 | (Hollander et al. 2001) |
| Blood density | 1060 kg/m$^3$ | |
| Dynamic viscosity | 0.0035 Pa sec | |

4.2. Symmetric Tree

An advantage of using a symmetric tree is that any vessel is a representative of all the vessels within the same generation. Thus, the results can be evaluated with respect to generations rather than individual vessels. The number of generations is equal to 19 and is evaluated by reaching the minimum branch radius (50μm).

### 4.2.1. Case 1: Constant Mass Fractions

Diameters of vessels as a result of the homeostatic optimization are shown **Fig. 4a**. The morphometry study (Huang et al. 1996) reported 15 orders of arteries in a human lung using the Strahler ordering system, where the 1$^{st}$ and 15$^{th}$ order vessels correspond to precapillary vessels and right/left pulmonary arteries, respectively. A comparison of our results with data from this study indicates an agreement between our first generation and their 14$^{th}$ order vessels, located two bifurcations downstream to the pulmonary trunk. In addition, the exponent $\xi$ in the daughter-to-parent radius relation (**Fig. 4c**) is approximately cubic along the tree (consistent with the original Murray's law). The daughter-to-parent area ratio $a_d/a_p$ is almost constant and falls in the range 1.2-1.3, associated with an open-end type of reflection (**Fig. 4d**), **Remark 2**. The vessel length, approximated as $L = 6.2R^{1.1}$ mm, is depicted in **Fig. 4b**, where it is compared to human data.

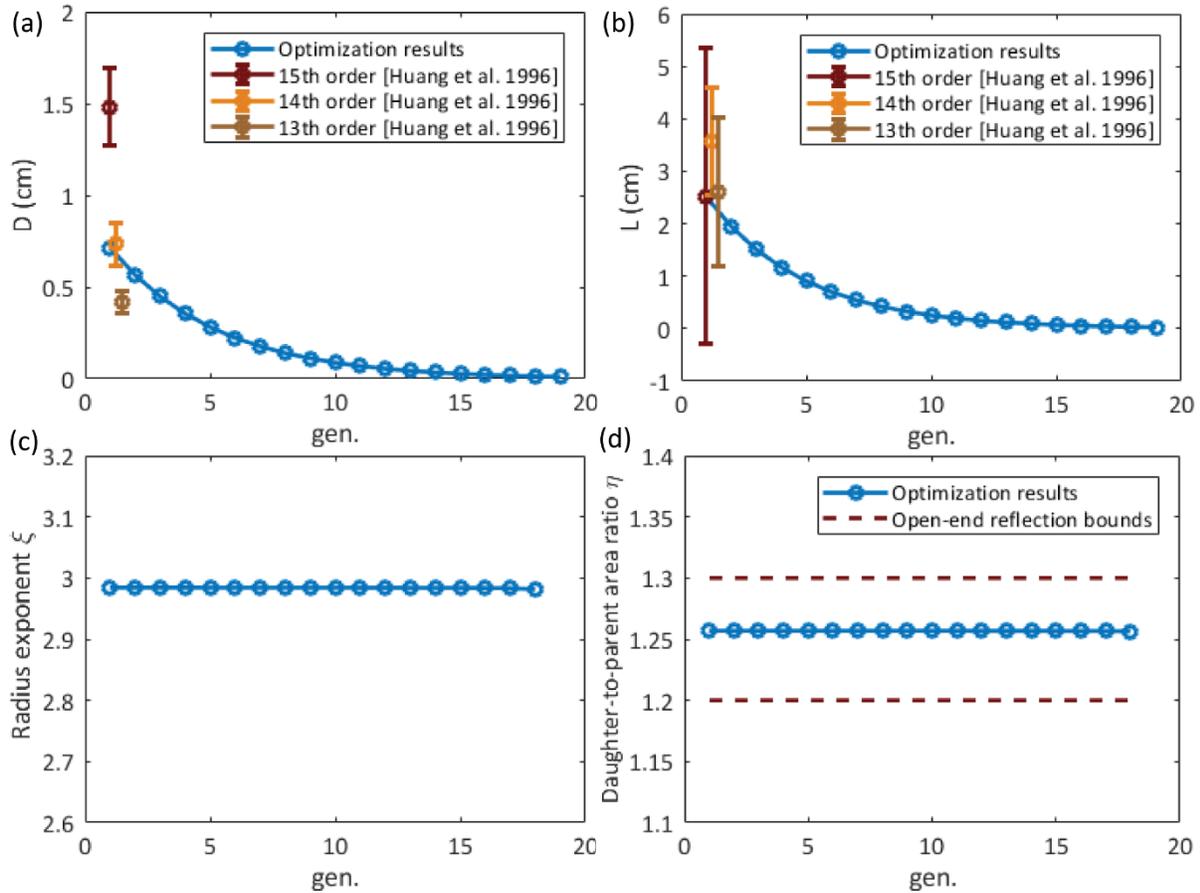

**Fig. 4** Symmetric tree – homeostatic optimization results plotted versus generation number: (a) diameter distribution compared to reported data of large vessels; (b) length distribution compared to reported data of large vessels; (c) radius exponent in daughter-to-parent radii relation; (d) daughter-to-parent area ratio within the range of open-end type of reflections

**Figure 5a** shows the thickness-to-diameter ratio, as a result of the homeostatic optimization, and a range of values reported in experimental studies (Li et al. 2012; Rol et al. 2017). The ratio is increasing distally consistent with a trend observed in arterial networks in the systemic circulation (Guo and Kassab 2004; Pries, Reglin, and Secomb 2005). Mid-pressure distribution along the tree is shown in **Fig. 5b**. The pressure gradient becomes steeper towards the terminal vessels where most of the pulmonary arterial resistance resides. Finally, homeostatic values of the mechanical stresses are obtained (**Fig. 5c,d**). The wall shear stress value $\tau$ is within 1-2.5Pa, observed in the systemic circulation (Kamiya, Bukhari, and Togawa 1984; Pries, Secomb, and Gaehtgens 1995). Noteworthy, in the original Murray's law, with $\xi = 3$, the wall shear stress, under Poiseuille law, is constant in the arterial tree (Zamir 1976). However, as $\xi$ deviates from 3, the wall shear stress increases (**Fig. 5c**) in agreement with measurements in the systemic microvascular networks. The decreasing trend of hoop stress (**Fig. 5d**) is also compatible with observations in the systemic circulation (Guo and Kassab 2004; Pries and Secomb 2005), although the values are significantly smaller.

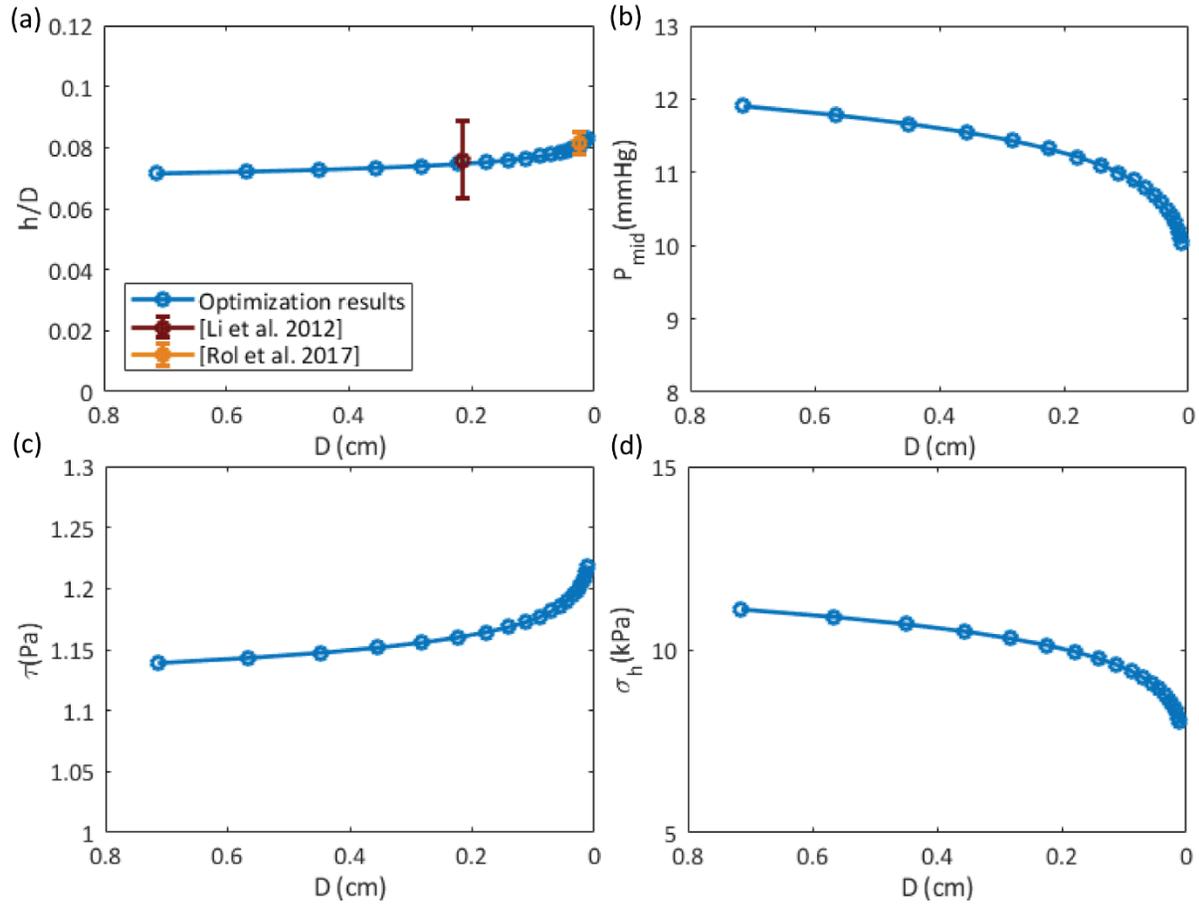

**Fig. 5** Symmetric tree – homeostatic optimization results plotted versus diameters: (a) wall thickness-to-diameter ratio; (b) mid-pressure; (c) homeostatic value of wall shear stress; (d) homeostatic value of circumferential stress

The results of the homeostatic optimization are linearized to obtain the orthotropic elastic membrane stiffness components: $E_{\theta\theta} = \left(\mathcal{A}_{\theta\theta}\mathcal{A}_{zz} - \mathcal{A}_{\theta z}^2\right)/\mathcal{A}_{zz}$, and $E_{zz} = \left(\mathcal{A}_{\theta\theta}\mathcal{A}_{zz} - \mathcal{A}_{\theta z}^2\right)/\mathcal{A}_{\theta\theta}$, in the arterial tree (**Fig. 6a**). The axial Young's moduli are markedly lower than those in the circumferential direction (Bernal et al. 2011; Baek et al. 2007) and both are decreasing across the generations.

It is an experimental challenge to estimate the arterial stiffness of distal pulmonary vasculature. Instead, a distensibility parameter $\lambda$ is introduced as $R/R_0 = 1 + \lambda p$, with pressure $p$ and radius at zero pressure $R_0$, to characterize the mechanical properties of arterial walls. A meta-analysis by Krenz and Dawson (Krenz and Dawson 2003) calculated a constant $\lambda^{KD} = 0.02$ mmHg$^{-1}$ by combining the experimental data across multiple species and different sizes of vessels. Moreover, a study on human specimens (Yen et al. 1990) reported $\lambda^{Yen} = 0.012$ mmHg$^{-1}$, almost constant along the pulmonary arterial tree. For an isotropic incompressible wall, the stiffness measure can be expressed via a distensibility parameter as $Eh/R_0 = 3/(4\lambda)$ (Qureshi et al. 2014), where $Eh$ is the structural stiffness (Humphrey et al. 2017). The homeostatic optimization yields a mildly decreasing ratio $E_{\theta\theta}h/R_0$ (**Fig. 6b**) that is near the constant value $\lambda^{Yen}$ and is above $\lambda^{KD}$. **Fig. 6b** includes the stiffness measure in the first generation estimated from porcine experimental data (Wang et al., n.d.).

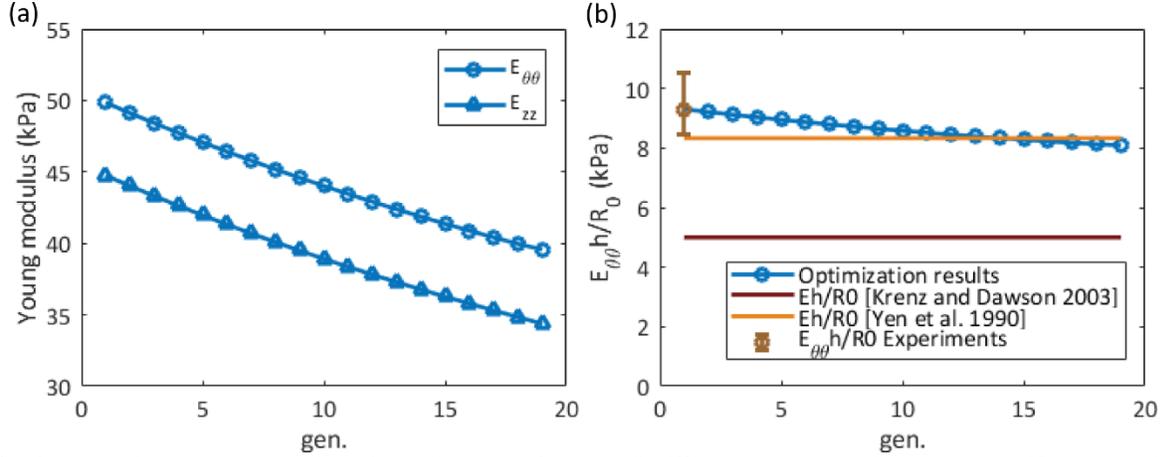

**Fig. 6** Symmetric tree – homeostatic optimization results for the wall stiffness versus generation number: (a) Young's modules in circumferential and longitudinal directions; (b) stiffness measure – circumferential structural stiffness divided by unstressed radius and compared to distensibility relations from data.

Next, we demonstrate the results of the fast-time hemodynamic analysis. The most valuable outcome from modeling the pulsatile flow in the distal vasculature is an estimation of the pulse wave velocity across the tree (**Fig. 7a**). The pulse wave velocity $c$ (6) depends on the wall stiffness as well as Womersley's number $\alpha^{Wom}$, which is shown for the 1$^{st}$ and the 9$^{th}$ harmonics (**Fig. 7b**). Thus, the steep decrease of the pulse wave velocity across the generations is the result of variability in vessel dimensions, decrease in $\alpha^{Wom}$, and heterogeneity of wall stiffness **Fig. 6b**. The Moens-Korteweg pulse wave velocity $c^{MK}$ is plotted for a comparison. The pulse wave velocity as a result of the homeostatic optimization agrees with the experimental data measured for large human pulmonary arteries (Banks et al. 1978; Milnor et al. 1969).

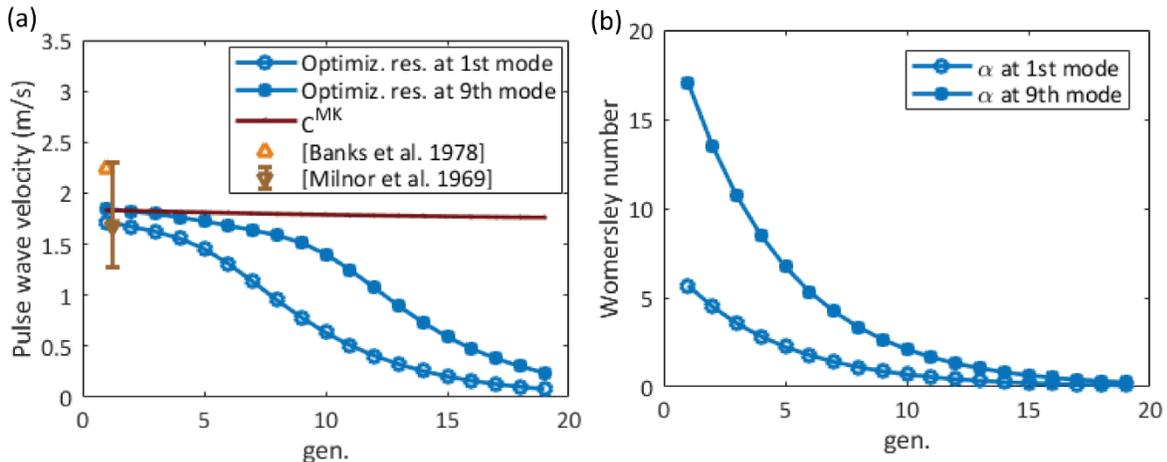

**Fig. 7** Symmetric tree – pulsatile hemodynamics results versus generation number: (a) pulse wave velocity for two harmonics, compared to data and Moens-Korteweg speed values; (b) Womersley's number for two harmonics.

Next, we justify the use of the deformable wall Womersley's theory (**Section 2.2**) by computing the ratio of maximum oscillatory velocity to the pulse wave speed $\delta_{\max} = \max_t(v_z^f/c)$, which must be small to guarantee the long wave approximation (Filonova et al. 2019). **Fig. 8** shows that there might be some contribution from the flow convection term in the first few generations, however, it is neglected in the current study, **Remark 5**.

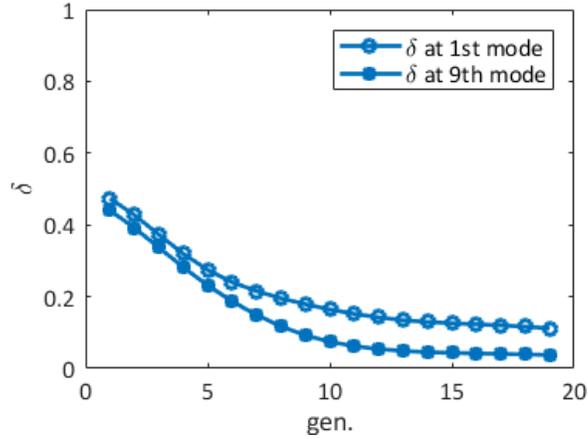

**Fig. 8** Symmetric tree – Womersley's solution validity check as function of generation number for two different harmonic modes.

The total hemodynamics solution (7) along the vascular tree is demonstrated in **Fig. 9**. The total input flow splits evenly at each generation. The total terminal pressure is within the physiological range of 8-25 mmHg. The minimum diastolic pressure falls below the mean pressure in proximal vessels, which can be explained by the negative type of wave reflection. Note that using equation (10) we obtain negative values of the reflection coefficients at the end of each vessel in the range of (-1,-0.54) considering real parts of the first mode values.

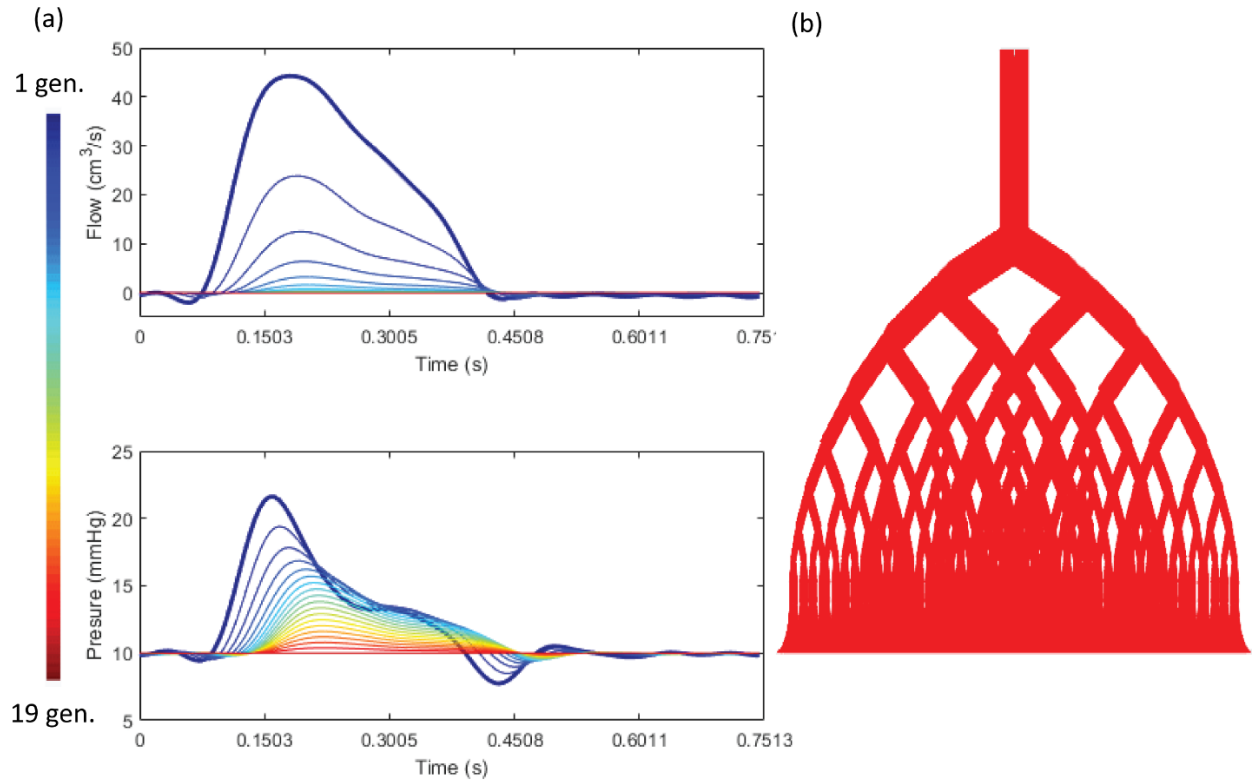

**Fig. 9** Symmetric tree: (a) total input flow and terminal pressure over the generations; (b) tree geometry (vessels radii and lengths) from homeostatic optimization

Noteworthy, the modeling framework presented in this study can facilitate further simulations of pulmonary circulation. For instance, the input impedance computed at the root vessel can be used as an outflow boundary condition for hemodynamic simulation in large vessels (Vignon-Clementel et al. 2006). **Fig. 10** shows the hemodynamics and all types of impedance (i.e. input, terminal, and characteristic impedance) at the root vessel of the modeled arterial tree (**Fig. 9b**).

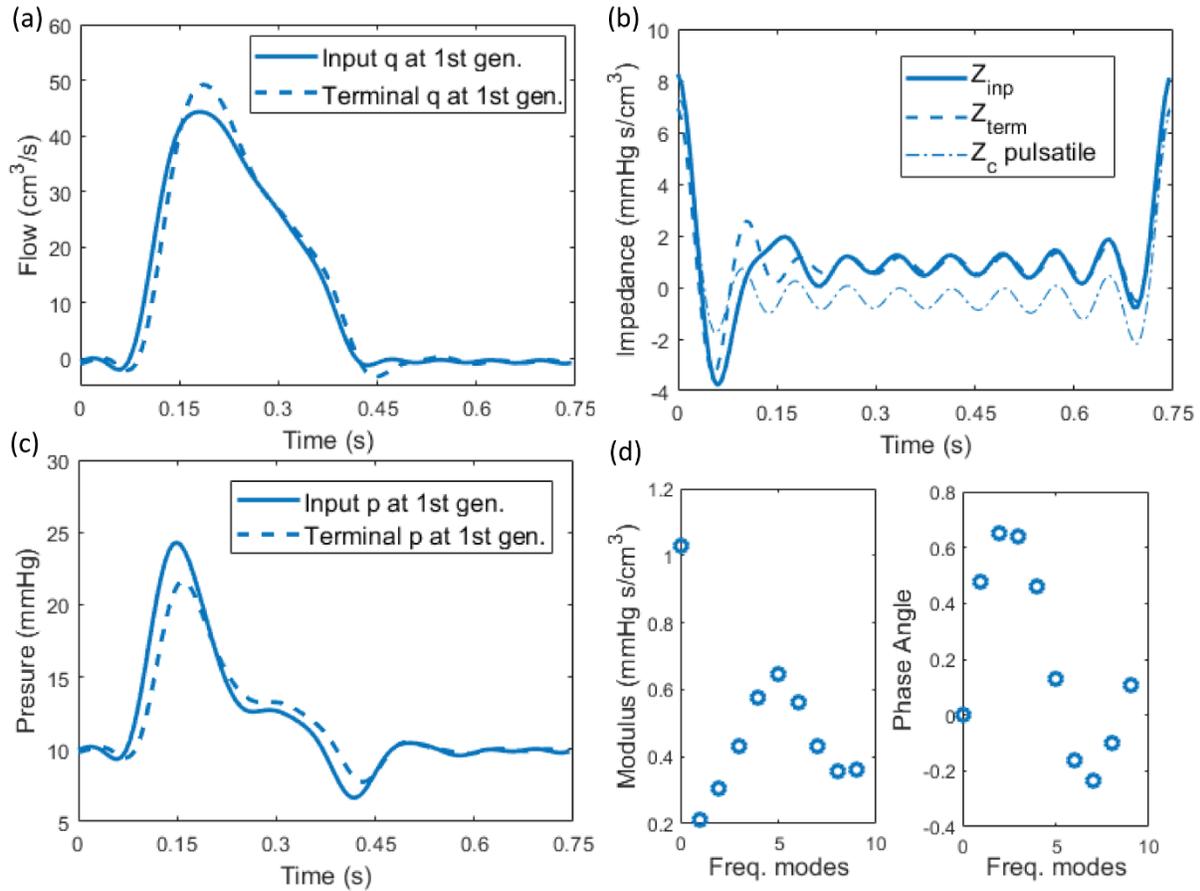

**Fig. 10** Symmetric tree: (a,c) total flow and pressure at the root vessel of the distal tree; (b) root vessel input impedance in time domain compared to terminal and characteristic impedance; (d) input impedance modulus and phase angles in frequency domain.

### 4.2.2. Case 2: Variable Mass Fractions

Arterial wall composition varies throughout the pulmonary arterial tree and, thus, mass fractions of vessel wall constituents are not constant (Hislop and Reid 1978). We take this into account by modifying mass fractions based on arteries composition (see **Section 4.1**), **Fig. 11**.

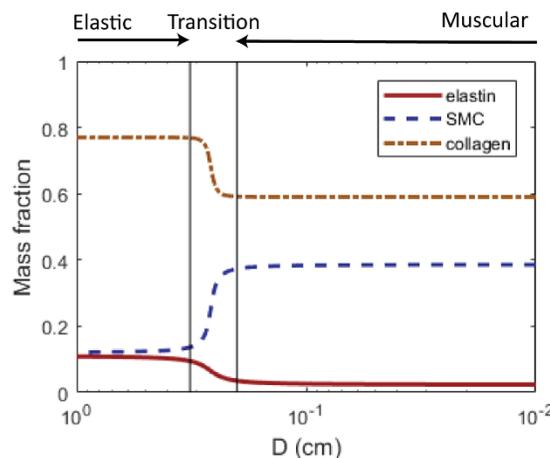

**Fig. 11** Prescribed variable mass fractions of the wall constituents: elastin, smooth muscle cells and collagen. The arrows on the top show the trend in arterial composition.

**Fig. 12** shows the homeostatic optimization results which significantly differ from the results of Case 1. During the transition region (**Fig. 11**), elastin is replaced by two materials that require metabolic energy for maintenance. The optimization, however, prefers a smaller vessel with more percentage of collagen and SMCs to a large vessel with less percentage of collagen and SMCs. This is reflected in a drop in the radius exponent (**Fig. 12a**). Moreover, the drop in $\xi$ results in a step-increase of the wall shear stress (**Fig. 12b**). **Fig. 12c** shows a step-drop in longitudinal Young's modulus caused by replacement of isotropic elastin matrix with SMCs that contribute to stiffness of the wall only in the circumferential direction. There is a slight step-increase in the stiffness measure $Eh/R_0$ leading to an increased average and a reduced variability along the tree **Fig. 12d**.

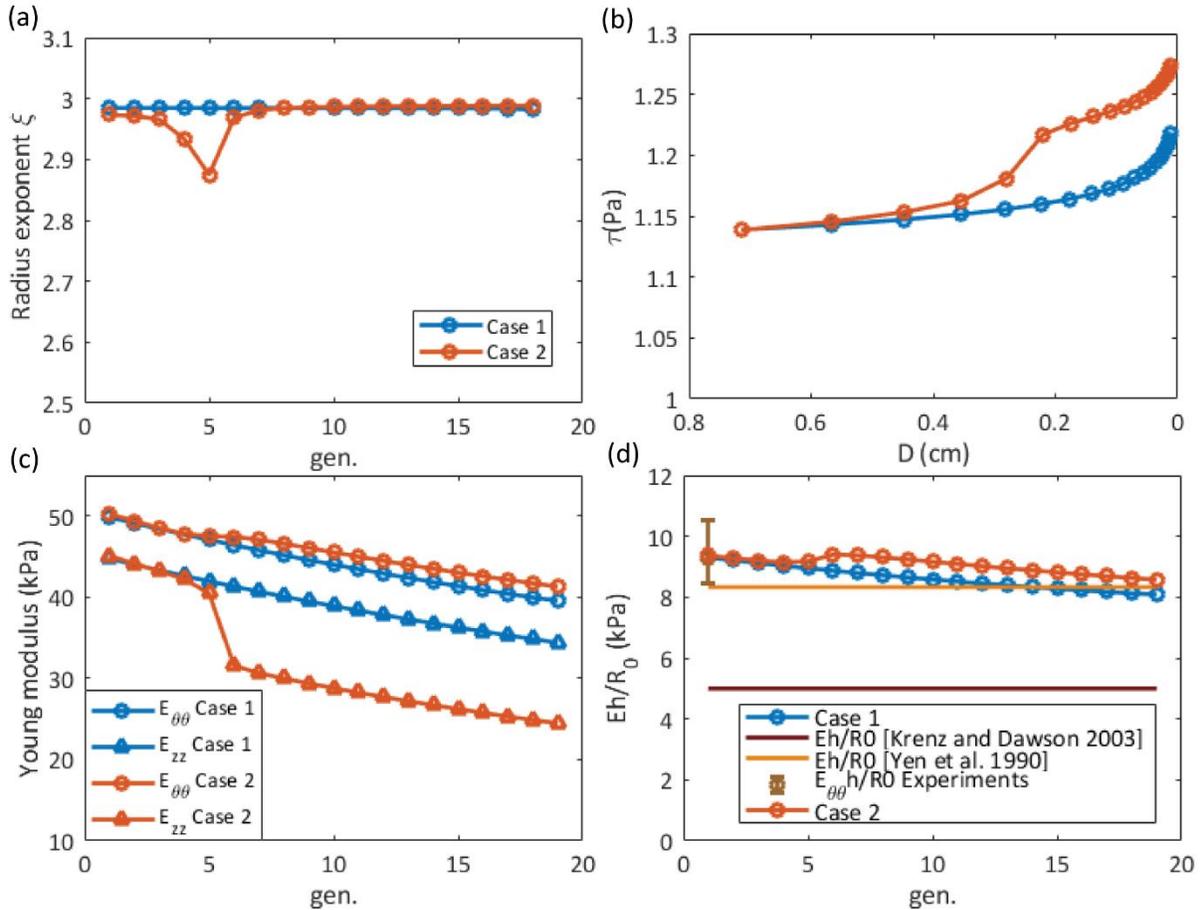

**Fig. 12** Symmetric tree – homeostatic optimization results of Case 1 (red) compared with Case 2 results (blue): (a) radius exponent in daughter-to-parent radii relation; (b) homeostatic value of wall shear stress; (c) Young's modulus in circumferential (top) and longitudinal (bottom) directions; (d) circumferential structural stiffness divided by unstressed radius

### 4.3. Asymmetric Tree

This section illustrates an example of the application of our homeostatic optimization and hemodynamic simulation in an asymmetric tree. The initial geometry of the tree is predefined by the parameters: $a_d/a_p$, $\xi$, $R_1$ and $R_{\min}$. The structure of the tree is initialized using a fractal rule with scaling factors $\alpha = 0.8$ and $\beta = 0.74$ defined from area ratio $a_d/a_p = 1.2$ and radii exponent $\xi = 2.72$. The root vessel and minimum radii are initialized as $R_1 = 0.36$ cm (from symmetric tree results) and $R_{min} = 0.018$ cm, respectively. A bifurcation is pruned if any of daughter radii is less than $R_{min}$. Thus, the initial tree had 15 generations and 32,767 vessels. After pruning the bifurcations, the number of generations and vessels were reduced to 14 and 4,233, respectively.

In this example, we use the parameters from **Table 1** (Case 2). For boundary conditions, however, the pressure is prescribed at the inlet and uniform flow is prescribed at outlets. The input pressure is taken from the symmetric tree results as 11.918 mmHg. The flow at outlets is prescribed by assuming an even split of input flow, 11.65 ml/s, among all outlets.

The homeostatic optimization modifies the tree geometry toward a more symmetric structure where over 60% of bifurcations are symmetric. This shows that a symmetric structure is more energy efficient. In addition, the optimization results in a uniform outlet radii to accommodate the prescribed uniform flow. **Fig. 13** shows the slow-time hemodynamic and structural results of the optimization. The blue and red lines indicate the longest and shortest paths along the arterial tree, respectively. All of the individual vessel segments are denoted by dots. The distributions of steady-state pressure and flow rate are illustrated for all the vessels in **Fig. 13a,b**. Clearly, the shortest and longest path results bound the distributions. Furthermore, the stratification of pressure along the generations shows a low variability of pressure within each generation, **Fig. 13b**. The stiffness measure and wall shear stress follow the same trends as Case 2 of the symmetric example. The stiffness measure is distributed around 9 kPa and is above the experimental estimations (**Fig. 13c**). In addition, the wall shear stress increases toward outlets and it has the same value for all outlets (**Fig. 13d**).

Next, we simulate the pulsatile hemodynamics in the asymmetric tree. To justify the application of the Womersley's theory in our asymmetric example, we compute $\delta_{max}$ (**Fig. 14a**). This parameter appears to be larger across the shortest paths in the arterial tree compared to the longest path. Our numerical experiments showed that the value of $\delta_{max}$ is even higher when a case of a highly asymmetric tree is considered. Therefore, we limited our example in this study to a mildly asymmetric tree ($\gamma = 0.86$). In addition, the pulse wave velocity varies from the shortest to the longest path in the arterial tree (**Fig. 14b**).

The hemodynamics in asymmetric tree are plotted for the longest and shortest paths in **Fig. 15d**. The total input flow splits at every bifurcation, giving more portion of flow to the longest path **Fig. 15a,c**. The total terminal pressure clearly has negative wave reflections. The impedances at the root vessel (**Fig. 15b**) differ from those in the symmetric case (**Fig. 10b**).

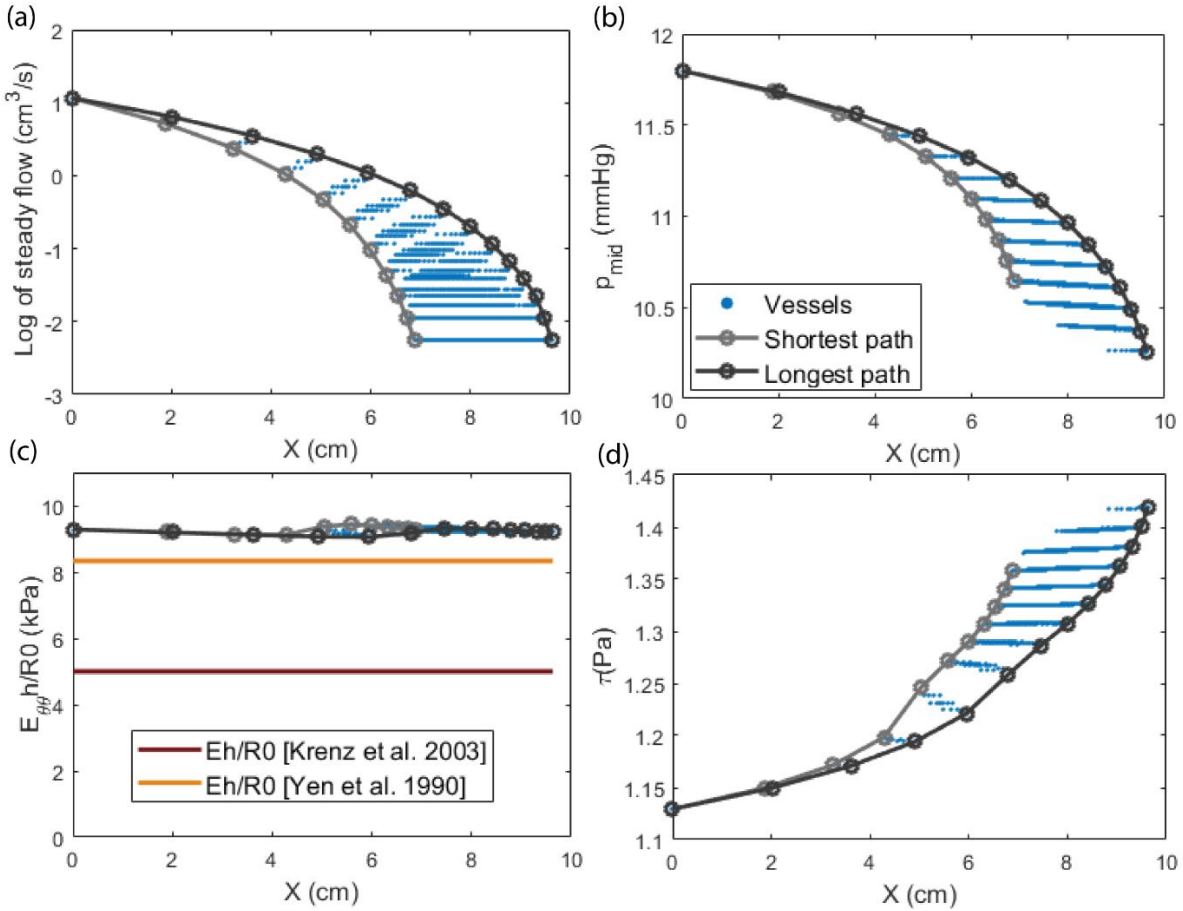

**Fig. 13** Asymmetric tree – homeostatic optimization results versus distance (from the root, along the branch pathway), dots represents optimization results for each vessel, red and blue lines indicate the short and long path, respectively: (a) log of steady-state flow rate; (b) terminal steady pressure; (c) ratio of structural stiffness to unstressed radius; (d) homeostatic value wall shear stress.

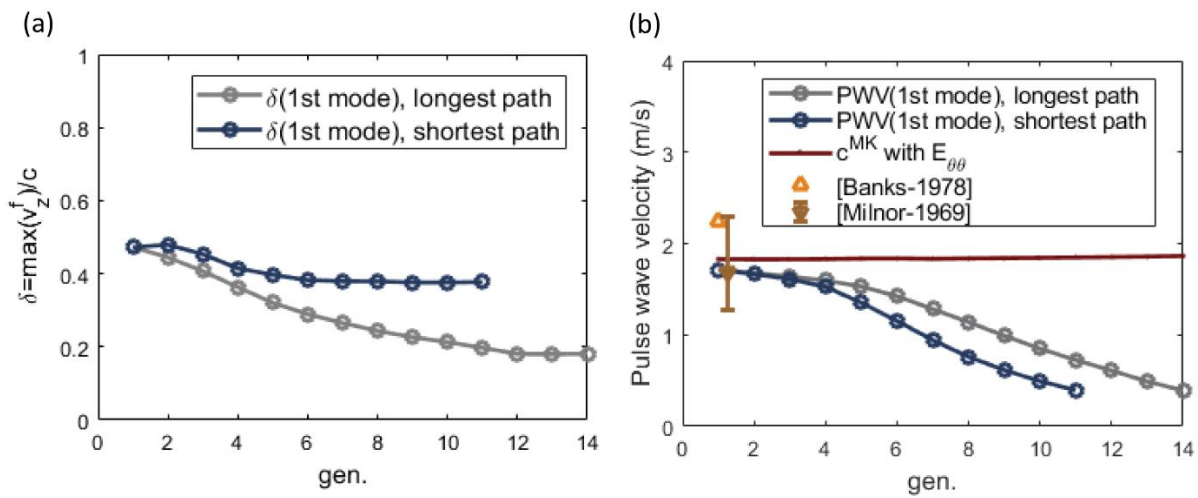

**Fig. 14** Asymmetric tree – pulse wave velocity and delta parameter for long and short pathways.

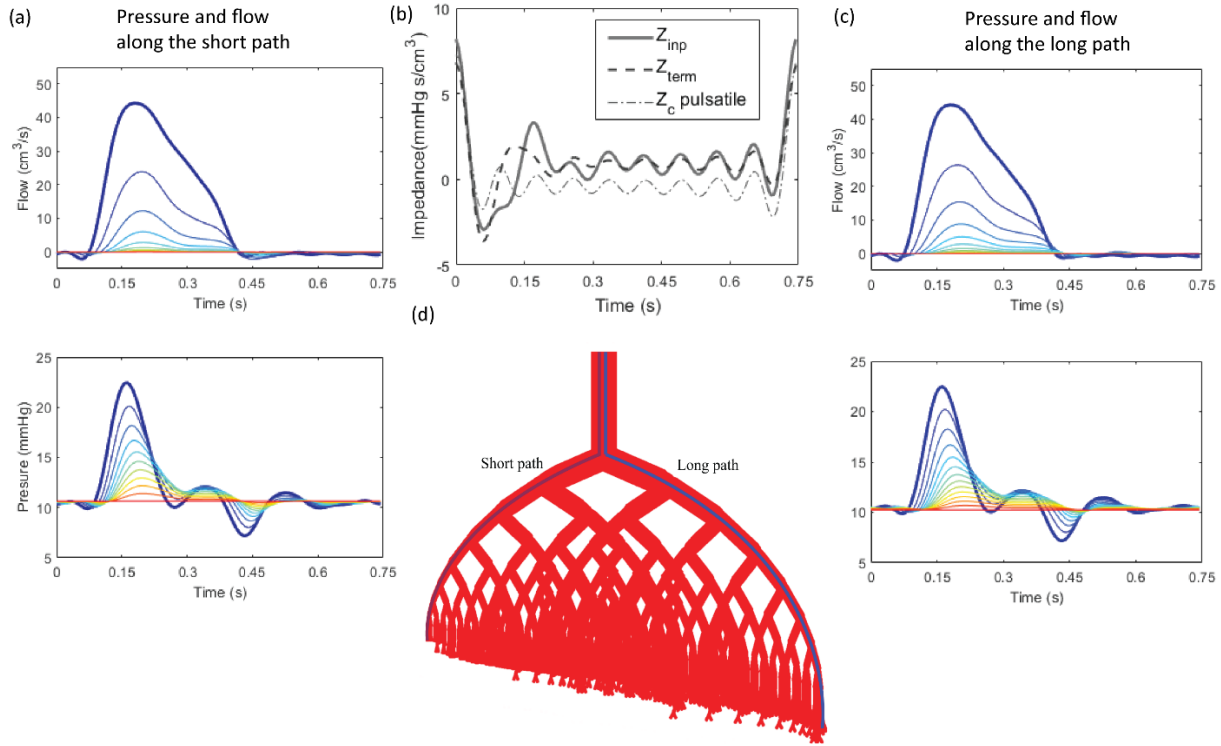

**Fig. 15** Asymmetric tree – the hemodynamics resulted: (a,c) total pressure and flow along the short and long pathways; (b) input, terminal and characteristic impedance at the root vessel; (d) tree geometry (vessels radii and lengths) from homeostatic optimization

## Discussion

*Summary:* Experimental studies have established the multiscale nature of PAH development and emphasized the importance of vascular remodeling in proximal and distal arteries. At the constitutive level, the vascular remodeling involves collagen deposition, degradation of elastin, and SMCs proliferation, which lead to the stiffening and thickening of the arterial wall. Hemodynamic factors in the cardiopulmonary system are major contributors to vascular G&R and these pathologies. Coupling between arterial structure modifications and hemodynamics have been established (Humphrey et al. 2009, 2017). In fact, vascular homeostasis is in equilibrium between hemodynamic forces and an arterial wall structure, characterized by wall thickness and stiffness, mechanical stresses, arterial tree configuration, etc. (Humphrey, Dufresne, and Schwartz 2014). Vascular remodeling is a departure from the homeostatic baseline. Therefore, to understand vascular remodeling in PAH, we need to define the homeostatic characteristics of a healthy pulmonary vasculature.

In this work, we successfully developed a novel method for establishing the homeostatic baseline for a pulmonary arterial network using metabolic energy considerations. In addition, we generalized a fluid-solid growth framework from a single vessel to an arterial tree, which can account for tissue mechanics (homeostatic state, G&R, etc.) and pulsatile hemodynamics. Based on an extended Murray's law, we assume that the equilibrium state of an arterial network minimizes the total metabolic energy consumption. Therefore, a cost function was formulated to incorporate the metabolic needs for blood supply in a vessel and for maintenance of the vessel wall constituents, and power dissipation due to blood viscosity. The optimization of this cost function resulted in the vessel radii, content of constituents, and steady state blood pressure and flow rates in the arterial tree. Through our FSG framework, the results of the homeostatic optimization were used to obtain the pulsatile hemodynamics using Womersley's analytical solution. Particularly, the orthotropic material behavior of the vessel wall was described by a nonlinear constitutive law in the homeostatic optimization, and then it was linearized around the steady pressure to compute the pulsatile solution.

In this work, we outlined the key ideas for developing a temporal multiscale framework to study G&R. This framework provides a one-way coupling between the slow- and fast-time scales, which significantly reduces the computational burden associated with fully coupled problems. Therefore, our framework is a computationally efficient tool that greatly facilitates the biomechanical analysis of vascular trees. Indeed, it allows for the focusing primarily on complex wall tissue processes and associated hemodynamic stimuli, under the consideration of the overall energetic efficiency in the vascular tree.

We illustrated the functionality of the framework on examples of symmetric and asymmetric fractal trees representing small- to medium-size vessels (0.01 to 0.5 cm) of the pulmonary vasculature. The homeostatic optimization provided the wall composition content, vessel wall stiffness, and homeostatic stresses, which are valuable intrinsic wall properties that otherwise are not measurable. The results for the stiffness measure ($Eh/R_0$) are close to the constant value related to experiments in (Yen et al. 1990): for the symmetric case, the stiffness measure slightly decreases toward the tree downstream and, for the asymmetric case, it varies near constant, which is slightly higher than the experimental value. The results for pulse wave velocity show good agreement with available experimental observations for the first few generations. The results for the thickness-to-diameter ratio at the middle and distal region of a tree are in a range of values reported in experimental studies.

Furthermore, the low computational cost of our model renders it immensely useful for parametric studies, as illustrated by considering cases of constant and variable mass fractions of wall constituents. This study revealed that mass fraction variability affects mostly radii exponent, wall shear stress, and Young's modulus distributions along the generations.

In addition, the hemodynamic solution gave the total pressure and flow distribution as well as the pulse wave velocity along the entire tree. Our results demonstrated a decrease in pulse wave velocity over the generations as a consequence of wall stiffness and Womersley's number variabilities. The proposed framework is also useful for obtaining the impedance outflow boundary conditions essential for coupling distal vasculature with large vessel 3D hemodynamic simulations in the patient-specific models. We calculated this input impedance for the symmetric and asymmetric trees.

*Limitations*: We observed that optimizing the pruned asymmetric fractal tree with the prescribed input flow and output terminal pressure leads to a trivial symmetric solution with a single prevailed bifurcation. To avoid this solution, we changed the boundary conditions to input pressure and output flow. In addition, for our examples, we considered the mild asymmetry of a tree to satisfy 'linearity' assumptions for Womersley's theory. However, modeling a highly asymmetric tree is possible using nonlinear hemodynamics theories.

Further work is required to improve and validate the method. The one-way coupling framework cannot account for the role of pulse pressure, which has been shown to be important during the progression of hypertension (Briones, Salaices, and Vila 2007). To address this limitation, the presented multiscale framework needs to be extended to include additional intermediate time scales and include physical regimes that are characterized by bi-directional coupling between governing equations at different time scales. However, we remark that contrary to the one-way coupling formulation, the full-coupling will require tremendous computational efforts due to nonlinearity, the need of using nested iterative loops, and large size of the problem.

In this paper, we used a constant blood viscosity, which is justified for the intermediate range of vessel size. However, for small vessels, such as pre-capillaries, the apparent viscosity should be considered, which depends on the vessel size and hematocrit level due to Fåhræus-Lindqvist effect (Pries and Secomb 2005). In this case, the total framework must be generalized to accommodate the viscosity variability.

Additional data and work are needed to improve the pulmonary arterial tree model. In presented examples, the tree is initialized as a bifurcating fractal structure based on a radial rule. With a few modifications, the proposed model can be applied to a vascular network that is reconstructed from an experimental-based morphometry of the pulmonary arterial tree.

Another limitation of the current study is the use of data from different species for the homeostatic optimization. Moreover, we only consider healthy subjects in our examples. Therefore, to improve the framework, there is a need for more experimental data on pulmonary arteries (especially small vessels) in healthy and PAH human subjects. Furthermore, a rigorous parameter sensitivity analysis and uncertainty quantification should be performed to enhance prediction capability of the framework.

## Acknowledgment
This work was supported by the National Institute of Health under Grant U01 HL135842.

## Appendix
    A. Temporal Multiscale Formulation

In this section, we outline a formal multiscale approach to consider coupling between the different temporal scales in G&R. Here, we present the key ideas behind the derivation of a multiscale formalism and illustrate a special case that allows us to obtain one-way coupled dynamics for the two scales.

*General Kinematics*

G&R is characterized by various phenomena that occur at different time scales: during the cardiac cycle, at the order of 1 second, SMC respond in the order of few minutes and vessel wall adaptation at the order of days. However, for the purpose of illustration of the key ideas of a formal temporal multiscale framework, we limit our study to two time scales only: a fast time $t$, characterizing the cardiac cycle, and a slow time $s$, characterizing the tissue adaptation process:

$$s = \varepsilon t, \tag{20}$$

where $\varepsilon \ll 1$ is a scale separation parameter. Physically, it can be understood as a scale that zooms-out the fast-time window interval, e.g. $t \in [0,1]$ sec., to its relatively small size at slow-time coordinates: $s \in [0, 1/86400]$ day, $\varepsilon = 1/86400$.

We use $\boldsymbol{X}$, $\boldsymbol{x}$ and $\boldsymbol{y}$ to denote the position in the reference configuration, intermediate configuration, and the current configuration respectively, **Fig. 16**. Next, we define two components of the total motion: a slow motion $\boldsymbol{u}^s(\boldsymbol{X},s) \sim O(1)$ and a fast perturbation motion $\eta \boldsymbol{u}^f(\boldsymbol{x},t) \sim O(\eta)$, where $\eta \ll 1$ is the kinematic parameter indicating the relative order of the slow and fast motion. We remark here that while we consider $\eta \ll 1$ for the purpose of this example, other modeling choices concerning $\eta$ as well as the time-scale separation parameter $\varepsilon$ exists and serve to define different physical regimes (Nama, Huang, and Costanzo 2017).

With the aforementioned choices for the temporal and spatial scales, the total motion is:

$$\boldsymbol{u}(\boldsymbol{y},t) = \boldsymbol{u}^s(\boldsymbol{X},s) + \eta \boldsymbol{u}^f(\boldsymbol{x},t), \quad \boldsymbol{x} = \boldsymbol{X} + \boldsymbol{u}^s, \quad \boldsymbol{y} = \boldsymbol{x} + \eta \boldsymbol{u}^f, \tag{21}$$

Hereafter, superscript $s$ stands for slow time and superscript $f$ stands for fast-time.

Next, we define the following deformation gradients corresponding to the fast, slow, and total motion:

$$\boldsymbol{F}^f = \frac{\partial \boldsymbol{y}}{\partial \boldsymbol{x}} = \boldsymbol{I} + \eta \boldsymbol{H}, \quad \boldsymbol{H} = \nabla_x \boldsymbol{u}^f, \tag{22}$$

$$\boldsymbol{F}^s = \frac{\partial \boldsymbol{x}}{\partial \boldsymbol{X}} = \boldsymbol{I} + \nabla_X \boldsymbol{u}^s, \quad \boldsymbol{F} = \frac{\partial \boldsymbol{y}}{\partial \boldsymbol{X}} = \boldsymbol{F}^f \boldsymbol{F}^s, \tag{23}$$

where $\boldsymbol{H} \sim O(1)$ is the fast-time displacement gradient.

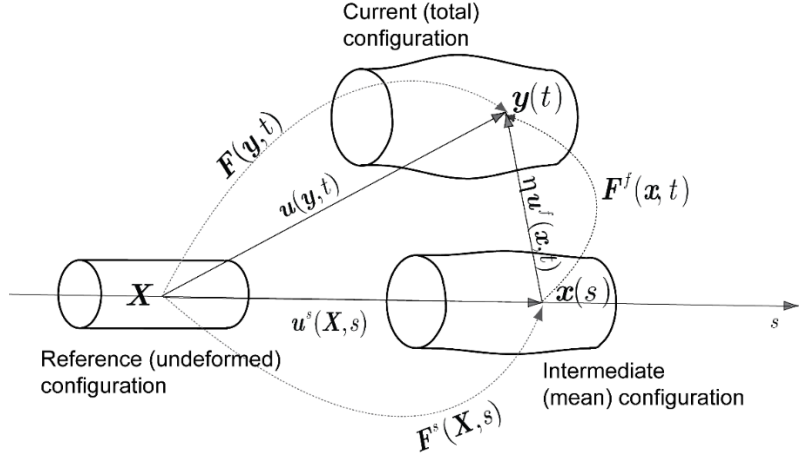

**Fig. 16** Configurations associated with two time scales: $s$ and $t$.

The velocity field is defined as the material time derivative of the displacement. From (20), (21), and (22), we have:

$$\boldsymbol{v}(\boldsymbol{y},t) = \left.\frac{\partial \boldsymbol{u}}{\partial t}\right|_{\boldsymbol{X}} = \varepsilon \boldsymbol{F}^f \boldsymbol{v}^s(\boldsymbol{X},s) + \eta \boldsymbol{v}^f(\boldsymbol{x},t) = \varepsilon \boldsymbol{v}^s + \eta \boldsymbol{v}^f + O(\varepsilon \eta), \qquad (24)$$

with slow and fast-time velocities, $\boldsymbol{v}^s, \boldsymbol{v}^f \sim O(1)$, defined as (Nama, Huang, and Costanzo 2017):

$$\boldsymbol{v}^f(\boldsymbol{x},t) = \left.\frac{\partial \boldsymbol{u}^f}{\partial t}\right|_{\boldsymbol{x}}, \quad \boldsymbol{v}^s(\boldsymbol{X},s) = \left.\frac{\partial \boldsymbol{u}^s}{\partial s}\right|_{\boldsymbol{X}}, \quad \left.\frac{\partial \boldsymbol{x}}{\partial s}\right|_{\boldsymbol{X}} = \varepsilon \boldsymbol{v}^s. \qquad (25)$$

It is convenient to consider the intermediate configuration as a configuration where the governing equations of the fluid and solid are mapped and solved (Figueroa et al. 2009). Following (Nama, Huang, and Costanzo 2017), the material derivative of the velocity mapped to the intermediate configuration is:

$$\left.\frac{d\boldsymbol{v}}{dt}\right|_{\boldsymbol{y}} = \left.\frac{\partial \boldsymbol{v}}{\partial t}\right|_{\boldsymbol{x}} + \left(\nabla_{\boldsymbol{x}} \boldsymbol{v}\right) \boldsymbol{F}^{f^{-1}}(\boldsymbol{v} - \eta \boldsymbol{v}^f) = \eta \left.\frac{\partial \boldsymbol{v}^f}{\partial t}\right|_{\boldsymbol{x}} + O(\varepsilon \eta), \qquad (26)$$

where $\boldsymbol{F}^{f^{-1}} = \boldsymbol{I} - \eta \boldsymbol{H}$. Next, we introduce an operator for time-averaging over the cardiac cycle $T$ as

$$\langle f(t) \rangle = \int_t^{t+T} f(\tau)/T \, d\tau.$$

To obtain system decoupling useful for practical applications, we make further **Assumptions**:

1. Fast-time displacement and velocity $\boldsymbol{u}^f$, $\boldsymbol{v}^f$ are periodic over the cardiac cycle;
2. Averaging of the fast displacement yields zero $\langle \boldsymbol{u}^f \rangle = 0$;
3. Higher-order terms $O(\varepsilon \eta)$, $O(\eta^2)$, $O(\varepsilon^2)$, etc. are similar and negligible, which is equivalent to assuming $O(\varepsilon) \sim O(\eta)$ and neglecting $O(\eta^2)$;
4. The pressure gradient is small and decomposed as: $\nabla_{\boldsymbol{x}} p = \varepsilon \nabla_{\boldsymbol{x}} p^s + \eta \nabla_{\boldsymbol{x}} p^f$ with $\langle \nabla_{\boldsymbol{x}} p^f \rangle = 0$.

**Remark 4.** Note that applying **Assumption 2** and time-averaging operator to (21) yields: $\langle \boldsymbol{u} \rangle = \boldsymbol{u}^s$. Also, using **Assumption 1** to averaging the fast-time velocity yields: $\langle \boldsymbol{v}^f \rangle \big|_{\boldsymbol{x}} = (\boldsymbol{u}^f(t+T) - \boldsymbol{u}^f(t))/T = 0$ and $\langle \partial \bm{v}^f / \partial t \rangle \big|_x = 0$. Thus, from (24) and (26) the mean velocity is equal to the slow-time velocity: $\langle \bm{v} \rangle = \varepsilon \bm{v}^s$ with negligible material derivative: $\langle d\bm{v}/dt \rangle \big|_y = O(\varepsilon \eta) \approx 0$. These help in deriving two-scales governing equations.

**Remark 5.** According to **Assumption 3** and (26), the inertia term is linear for linear momentum equations at the intermediate configuration, while its mean value is negligible (**Remark 4**). We note that if the finite wall fast-time deformations are considered ($\eta$ is no longer small) then the mixed fast-slow-time nonlinear term $(\nabla_x \bm{v}^f) \bm{v}^s$ can present in the system and be of order $O(\varepsilon)$. Indeed, alternative modeling assumptions concerning $\eta$ and $\varepsilon$ need to be employed to obtain different kinematic descriptions and physical regimes of the system. However, for the purpose of the current example, we limit ourselves to the cases where $\eta \ll 1$.

**Remark 6.** The kinematic modeling assumptions in this example consider that fast-time displacement is much smaller than that of slow-time (21), while fast-time velocity is comparable to that of slow-time (24), and thus the total velocity is of order smaller than slow-time displacement. These kinematics can be further extended to the fluid domain within the lumen by assuming pressure gradient of order consistent with the velocity field (24), as proposed in **Assumption 4**. Physically this can be representative of the regions with slow (attenuated) yet pulsatile blood flow, as in intermediate and distal parts of the vascular bed.

*Fluid governing equations*

Consider an incompressible Newtonian fluid with constant viscosity $\mu$ and density $\rho_{fluid}$. The balance of mass and balance of linear momentum (in absence of body force) mapped to the intermediate configuration (**Fig. 16**) yield (Nama, Huang, and Costanzo 2017):

$$\bm{F}^{f^{-T}} : \nabla_x \bm{v} = 0, \qquad \rho_{fluid} \frac{d\bm{v}}{dt}\bigg|_y = \nabla_x \cdot \bm{P}_{fluid}, \tag{27}$$

with the first Piola-Kirchoff stress tensor $\bm{P}_{fluid} = \bm{\sigma}_{fluid} \bm{F}^{f^{-T}}$ and Cauchy stress $\bm{\sigma}_{fluid} = -p\bm{I} + \mu(\nabla_x \bm{v} + \nabla_x \bm{v}^T)$. Applying (26) and **Assumption 4** we have:

$$\nabla_x \cdot (\varepsilon \bm{v}^s + \eta \bm{v}^f) = 0, \qquad \rho_{fluid} \eta \frac{\partial \bm{v}^f}{\partial t}\bigg|_x = -\varepsilon \nabla_x p^s - \eta \nabla_x p^f + \mu \varepsilon \nabla_x^2 \bm{v}^s + \mu \eta \nabla_x^2 \bm{v}^f. \tag{28}$$

The time-scale separation of governing equations can be obtained as follows. Averaging (28) over the cardiac cycle provides the *fluid slow-time governing equations*:

$$\nabla_x \cdot \varepsilon \bm{v}^s = 0, \qquad 0 = -\varepsilon \nabla_x p^s + \mu \varepsilon \nabla_x^2 \bm{v}^s, \tag{29}$$

with $\langle \bm{\sigma}_{fluid} \rangle = \varepsilon \bm{\sigma}_{fluid}^s = -\varepsilon \nabla_x p^s + \mu \varepsilon \nabla_x^2 \bm{v}^s$ and use of **Remark 4** and **Assumption 4**. Subtracting (29) from (27) provides the *fluid fast-time governing equations*:

$$\nabla_x \cdot \eta \bm{v}^f = 0, \qquad \rho_{fluid} \eta \frac{\partial \bm{v}^f}{\partial t}\bigg|_x = -\eta \nabla_x p^f + \mu \eta \nabla_x^2 \bm{v}^f. \tag{30}$$

*Solid Governing Equations*

The balance of linear momentum for the solid (deformable vessel wall) mapped onto the mean configuration can be written as:

$$\rho_{solid} \left.\frac{\partial^2 \boldsymbol{u}}{\partial t^2}\right|_{X} = \nabla_{\boldsymbol{x}} \cdot \boldsymbol{P}_{solid} + \mathbf{B}, \tag{31}$$

with body force $\mathbf{B}$, the first Piola-Kirchhoff stress tensor $\boldsymbol{P}_{solid} = \boldsymbol{\sigma}_{solid} \boldsymbol{F}^{f^{-T}}$, and Cauchy stress tensor $\boldsymbol{\sigma}_{solid}$. The vessel wall is described as a hyperelastic incompressible material (Baek et al. 2007; Figueroa et al. 2009):

$$\boldsymbol{\sigma}_{solid} = -p_L \boldsymbol{I} + \boldsymbol{F} \boldsymbol{S} \boldsymbol{F}^T, \qquad \boldsymbol{S} = 2\frac{\partial W}{\partial \boldsymbol{C}}, \tag{32}$$

with Lagrange multiplier $p_L$ to enforce the wall incompressibility; the right Cauchy-Green deformation tensor $\boldsymbol{C}$; the stored energy function $W(\boldsymbol{C})$; and a deviatoric part of the second Piola-Kirchhoff stress tensor $\boldsymbol{S}$. The latter can be linearized using a Taylor series expansion around the slow-time deformation measure $\boldsymbol{C}^s = \boldsymbol{F}^{sT} \boldsymbol{F}^s$ as shown in (Baek et al. 2007):

$$\boldsymbol{S}(\boldsymbol{C}) = \boldsymbol{S}^s + \eta \boldsymbol{S}^f + O(\eta^2), \quad \boldsymbol{S}^s = 2\left.\frac{\partial W}{\partial \boldsymbol{C}}\right|_{\boldsymbol{C}^s}, \quad \boldsymbol{S}^f = \left.\frac{\partial \boldsymbol{S}}{\partial \boldsymbol{C}}\right|_{\boldsymbol{C}^s} \boldsymbol{F}^{sT}\left(\boldsymbol{H} + \boldsymbol{H}^T\right) \boldsymbol{F}^s. \tag{33}$$

Substituting (22), (23) to the first Piola-Kirchhoff stress tensor yields:

$$\boldsymbol{P}_{solid} = \boldsymbol{\sigma}_{solid} \boldsymbol{F}^{f^{-T}} = \left(-p_L \boldsymbol{I} + (\boldsymbol{I} + \eta \boldsymbol{H}) \boldsymbol{F}^s (\boldsymbol{S}^s + \eta \boldsymbol{S}^f) \boldsymbol{F}^{sT} (\boldsymbol{I} + \eta \boldsymbol{H}^T)\right)\left(\boldsymbol{I} - \eta \boldsymbol{H}^T\right). \tag{34}$$

Assuming $p_L = \varepsilon p_L^s + \eta p_L^f$ and neglecting $O(\eta^2)$ in (34), we linearize the wall stress $\boldsymbol{P}_{solid}$ around the slow-time state as shown in (Figueroa et al. 2009):

$$\begin{aligned}
\boldsymbol{P}_{solid} &= \boldsymbol{P}_{solid}^s + \eta \boldsymbol{P}_{solid}^f, \quad \boldsymbol{P}_{solid}^s = -\eta p_L^s \boldsymbol{I} + \hat{\boldsymbol{P}}_{solid}^s, \quad \hat{\boldsymbol{P}}_{solid}^s = \boldsymbol{F}^s \boldsymbol{S}^s \boldsymbol{F}^{sT}, \\
\boldsymbol{P}_{solid}^f &= -p_L^f \boldsymbol{I} + \boldsymbol{H} \hat{\boldsymbol{P}}_{solid}^s + \boldsymbol{F}^s \boldsymbol{S}^f \boldsymbol{F}^{sT}.
\end{aligned} \tag{35}$$

For modeling the mechanical response of the vessels wall at the intermediate configuration, we need to linearize the material stiffness based on $\boldsymbol{P}_{solid}^s$ and $\left.\partial^2 W / \partial \boldsymbol{C}^2\right|_{\boldsymbol{C}^s}$. Thus, we introduce the symmetric stiffness tensor $\left.\mathcal{A}\right|_s$ that linearizes the hyperelastic response at a particular intermediate configuration. Then, the fast-time stress tensor at the particular intermediate configuration $\left.\boldsymbol{P}_{solid}^f\right|_s$ is expressed as follows (Figueroa et al. 2009):

$$\left.\eta \boldsymbol{P}_{solid}^f\right|_s = \left.\mathcal{A}\right|_s \frac{1}{2}\eta\left(\boldsymbol{H} + \boldsymbol{H}^T\right), \quad \left.\mathcal{A}_{ijkl}\right|_s \approx \delta_{ik}(P_{solid}^s)_{lj} + 4 F_{iA}^s F_{jB}^s F_{kP}^s F_{lQ}^s \left.\frac{\partial^2 W}{\partial C_{AB} \partial C_{PQ}}\right|_{\boldsymbol{C}^s}, \tag{36}$$

with infinitesimal strains and neglected rotations. Following the constrained mixture theory, the strain energy of the mixture, $W(\boldsymbol{C})$, includes the masses of the wall constituents and can incorporate the vascular G&R processes (Baek et al. 2007).

Finally, applying time-averaging operator to (31) we obtain *the solid slow-time governing equation*:

$$0 = \nabla_{\boldsymbol{x}} \cdot \boldsymbol{P}_{solid}^s + \mathbf{B}^s, \tag{37}$$

where mean stress $\langle \boldsymbol{P}_{solid} \rangle = \boldsymbol{P}_{solid}^s$, with assumed $\langle p_L^f \rangle = 0$; and mean body force $\mathbf{B}^s = \langle \mathbf{B} \rangle$ for decomposition $\mathbf{B} = \mathbf{B}^s + \eta \mathbf{B}^f$. Then subtracting (37) from (31) yields *the solid fast-time governing equation*:

$$\rho_{solid} \left.\frac{\partial^2 \boldsymbol{u}^f}{\partial t^2}\right|_{\boldsymbol{x}} = \left.\nabla_{\boldsymbol{x}} \cdot \boldsymbol{P}^f_{solid}\right|_s + \mathbf{B}^f. \tag{38}$$

For simplicity, in the manuscript, the variable scale factors $\varepsilon$ and $\eta$ are omitted as these factors can be embedded back: $p^s = \varepsilon p^s$, $p^f = \eta p^f$, etc.

### B. Slow-time Hemodynamics

Here we present an algorithm that computes the slow-time pressure and flow at each vessel for a given geometry of a bifurcating tree. We introduce the index $[k,m]$ to label an individual vessel within the tree generation $k$ for $m = 1,..2^{k-1}$. The large sparse matrix can be constructed to account for the pressure continuity and flow conservation at each bifurcation as well as Poiseuille flow resistance at each vessel. The block-matrix, related to the vessel $[k,m]$, comprises of flow split and Poiseuille equations (3) for daughter vessels in generation $k+1$ :

**Given resistance** $res[k,m]$ :

**Construct** $[k,m]$ **- block of equations :**

$$\begin{aligned} q^s[k+1,2m-1] + q^s[k+1,2m] &= q^s[k,\mathrm{m}] \\ p^s_T[k,\mathrm{m}] - p^s_T[k+1,2m-1] &= res[k+1,2m-1]q^s[k+1,2m-1] \\ p^s_T[k,\mathrm{m}] - p^s_T[k+1,2m] &= res[k+1,2m]q^s[k+1,2m] \end{aligned} \tag{39}$$

**for unknowns :**

$q^s[k,\mathrm{m}], q^s[k+1,2m-1], q^s[k+1,2m], p^s_T[k,\mathrm{m}], p^s_T[k+1,2m-1], p^s_T[k+1,2m]$

For boundary conditions in the slow-time hemodynamic system in a tree, we consider two options: (1) given input steady flow at the root vessel and terminal steady pressure at outlets; and (2) given input steady pressure at root vessel and steady flow at outlets. In the manuscript, for the numerical example of the symmetric tree, we consider boundary conditions (1) while for the asymmetric tree example (2).

For the unpruned tree with $N$ number of generations, the total number of equations (and unknowns) in the system is $2(2^N - 1)$ with total number of individual vessels $2^N - 1$. Taking into account boundary conditions by excluding $2^{N-1} + 1$ unknowns reduces number of unknowns to $3(2^{N-1} - 1)$.

### C. Fast-time Hemodynamics

Here we present an algorithm that for a given geometry of a bifurcating tree and discrete frequency $\omega_n$, computes the impedance and then fast-time hemodynamics at each vessel. First, using (10), (13), (14), we compute the impedance functions at each vessel from bottom-to-top of the tree. We consider that the terminal impedance or, equivalently, the reflection coefficient (10) at outlets is given:

**Given :** $\rho_{fluid}, \mu, h, T, \omega_n,$
$R[k,m], L[k,m], \mathcal{A}_{\theta\theta}[k,m] = \mathcal{A}_{\theta\theta}\big|_{p^s[k,m]}$ **for** $m = 1,..2^{k-1},$
$Z_n^T[N-1,m] = Z_n^{inp}[N,m]$ **for** $m = 1,..2^{N-1}$

**Find :** $Z_n^{inp}[k,m]$ **for** $k < N$ **and** $m = 1,..2^{k-1}$

for $k = (N-1) : -1 : 1$
   for m $= 1 : 2^{k-1}$

$$(g_n, c_n)[k,m] = \text{Womersley Solution}\, (\rho_{fluid}, \mu, h, T, R[k,m], \mathcal{A}_{\theta\theta}[k,m])$$

$$Z_n^T[k,m] = \frac{Z_n^{inp}[k+1, 2m] \cdot Z_n^{inp}[k+1, 2m-1]}{Z_n^{inp}[k+1, 2m] + Z_n^{inp}[k+1, 2m-1]}$$

$$Z_n^c[k,m] = \frac{\rho_{fluid} c_n[k,m]}{\pi R^2[k,m]\big(1 - g_n[k,m]\big)}$$

$$\Gamma_n[k,m] = \frac{Z_n^T[k,m] - Z_n^c[k,m]}{Z_n^T[k,m] + Z_n^c[k,m]}$$

$$Z_n^{inp}[k,m] = Z_n^c[k,m] \frac{1 + \Gamma_n[k,m] e^{-i\omega_n 2 L[k,m]/c_n[k,m]}}{1 - \Gamma_n[k,m] e^{-i\omega_n 2 L[k,m]/c_n[k,m]}} \tag{40}$$

   end
 end

Second, we reconstruct the pulsatile hemodynamics in the system by computing the pressure and flow solution at each vessel from top-to-bottom of the tree. The components of the fast-time terminal pressure $P_n^T$ and input flow $Q_n^{inp}$ in frequency domain are found using (12). For illustration of the algorithm, we consider the boundary condition at the root vessel as a given input pressure.

**Given :** $P_n^{inp}[1,1]$
$R, L, \omega_n, c_n, \Gamma_n, Z_n^c, Z_n^{inp}$ **at vessel** $[k,m]$

**Find :** $P_n^T[k,m], Q_n^{inp}[k,m]$

for $k = 1 : N$
   for $m = 1 : 2^{k-1}$

$$\bar{m} = \text{round}(m/2)$$
$$P_n^{inp}[k,m] = P_n^T[k-1, \bar{m}] \quad (\text{for } k > 1)$$
$$Q_n^{inp}[k,m] = P_n^{inp}[k,m] / Z_n^{inp}[k,m]$$
$$H_{forw} = P_n^{inp}[k,m] / \Big(1 + \Gamma_n[k,m] e^{-i 2\omega_n L[k,m]/c_n[k,m]}\Big)$$
$$P_n^T[k,m] = H_{forw} e^{-i\omega_n L/c_n[k,m]} \Big(1 + \Gamma_n[k,m]\Big)$$
\tag{41}

   end
 end

### D. Wall Mechanics in a Vessel

A single vessel of the arterial tree is considered as a thin-walled cylindrical tube composed of three main constituents: elastin, collagen, and SMC. Each constituent is assumed to separately contribute to the strain energy density per unit area ($w = \rho_{wall} W / M_R^{total}$): $w = w^{el} + w^{smc} + w^{col}$.

There is a separate prestretch mapping from each constituent natural configuration to the overall reference configuration (Figueroa et al. 2009). For instance, the prestretch mapping for elastin can be expressed as:

$$\mathbf{G}^{el} = \begin{bmatrix} G_r^{el} & 0 & 0 \\ 0 & G_\theta^{el} & 0 \\ 0 & 0 & G_z^{el} \end{bmatrix} \quad (42)$$

where $G_\theta^{el}, G_z^{el}$ are prestretches associated with circumferential and axial directions, and $G_r^{el} = \left(G_\theta^{el} G_z^{el}\right)^{-1}$.

Similarly, the collagen fibers and SMCs are continuously produced at constant homeostatic stretches along certain directions. Thus, we define $\mathbf{M}^i$ for $i \in \{k, smc\}$ as the unit vector in the direction of the $k$-th collagen fiber family or SMC. Then, prestretch mappings for collagen and SMC are given by $\mathbf{G}^k = G_h^{col} \mathbf{M}^k \otimes \mathbf{M}^k$, $\mathbf{G}^{smc} = G_h^{smc} \mathbf{M}^{smc} \otimes \mathbf{M}^{smc}$, where the prestretch factors $G_h^{col}$ and $G_h^{smc}$ are identified as constant homeostatic stretches, i.e., the stretches of the constituents when they are produced (Baek, Rajagopal, and Humphrey 2006). The orientation of collagen fibers and SMC in the reference configuration, defined by angle $\gamma^k$, can be written as:

$$\mathbf{M}^k = \cos\gamma^k \mathbf{e}_Z + \sin\gamma^k \mathbf{e}_\theta, \quad \mathbf{M}^{smc} = \mathbf{e}_\theta. \quad (43)$$

Next, for modeling an extension and inflation of a thin-wall vessel, we consider a deformation gradient at the slow-time scale: $\mathbf{F}^s = \mathrm{diag}[\lambda_r, \lambda_\theta, \lambda_z]$. The incompressibility of the wall material is imposed by assuming an isochoric motion, $\det \mathbf{F}^s = 1$, therefore, $\lambda_r = \left(\lambda_\theta \lambda_z\right)^{-1}$. Using membrane theory (Humphrey 2002), the membrane Cauchy stress as the force per deformed length can be defined by:

$$\mathbf{T} = \frac{1}{J_{2D}} \mathbf{F}^s \frac{\partial w}{\partial \mathbf{F}^{sT}}, \quad T_{\theta\theta} = \frac{1}{\lambda_z} \frac{\partial w}{\partial \lambda_\theta}, \quad T_{zz} = \frac{1}{\lambda_\theta} \frac{\partial w}{\partial \lambda_z}, \quad (44)$$

where $J_{2D} = \lambda_\theta \lambda_z$.

We use the following strain energy density functions for the three materials:

$$\begin{aligned}
\Psi^{el} &= \frac{c_1}{2}\left((\lambda_\theta^{el})^2 + (\lambda_z^{el})^2 + \frac{1}{(\lambda_\theta^{el})^2 (\lambda_z^{el})^2} - 3\right), \\
\Psi^k &= \frac{c_2}{4c_3}\left(e^{c_3\left((\lambda^k)^2 - 1\right)^2} - 1\right), \\
\Psi^{smc} &= \frac{c_4}{4c_5}\left(e^{c_5\left((\lambda^{smc})^2 - 1\right)^2} - 1\right) + \frac{S}{\rho_{solid}}\left(\lambda_\theta + \frac{1}{3}\frac{(\lambda_M - \tilde{\lambda})^3}{(\lambda_M - \lambda_0)^2}\right).
\end{aligned} \quad (45)$$

where $S$ is the basal active tone and $\tilde{\lambda}$ is the active stretch of the SMCs in the circumferential direction. In this study, we do not consider the long-term adaptations of the arteries, and thus, we assume $\tilde{\lambda} = \lambda_\theta$. The strain energy per unit area is

$$w = M_R^{el} \Psi^{el} + \sum_k M_R^k \Psi^k + M_R^{smc} \Psi^{smc}. \quad (46)$$

The stretches in each constituent $\lambda^i$ is expressed in terms of the prestretches using $\mathbf{F}^i = \mathbf{F}^s \mathbf{G}^i$:

$$\lambda_\theta^{el} = G_\theta^{el} \lambda_\theta, \quad \lambda_z^{el} = G_z^{el} \lambda_z, \quad \lambda^k = G_h^{col}\sqrt{\lambda_\theta^2 \sin^2\gamma^k + \lambda_z^2 \cos^2\gamma^k}, \quad \lambda^{smc} = G_h^{smc}\lambda_\theta. \quad (47)$$

Therefore, the circumferential membrane Cauchy stress at slow-time can be written as:

$$T_{\theta\theta}\big|_s = \frac{1}{\lambda_z}\left(M_R^{el}\frac{\partial \Psi^{el}}{\partial \lambda_\theta} + \sum_k M_R^k \frac{\partial \Psi^k}{\partial \lambda_\theta} + M_R^{smc}\frac{\partial \Psi^{smc}}{\partial \lambda_\theta}\right). \tag{48}$$

Finally, for pressurized thin-walled cylinder with pressure $\bar{p}^s$, the force equilibrium in the circumferential direction gives the Laplace law expression (4) as $2LT_{\theta\theta} = 2\bar{p}^s RL$, yielding

$$T_{\theta\theta}\big|_s = \bar{p}^s R , \tag{49}$$

where $R$ and $L$ are the artery deformed radius and length, respectively. Note that the reference configuration, $F = I$, pertains to a loaded configuration, where due to the prestretches, the strain energies of the constituents and their derivatives do not vanish. The homeostatic reference configuration is defined at the reference configuration (with prestretches) and the stresses associated with the constituents in this configuration are called homeostatic stresses ($\sigma^i$).

### E. Cost Function Minimization

The cost function minimization (19) at each individual vessel $[k, m]$ is solved together with the steady-state hemodynamics constrains acting at the entire tree (**Appendix B**). This minimization is implemented in the nested loop. In the external loop, given a current tree geometry the tree hemodynamics $p^s$, $q^s$ is resolved with use of (39). Then in the internal loop, given hemodynamic state, the optimization problem $C(R; \bar{p}^s, q^s)$ is solved sequentially at each vessel using the Newton-Raphson method to update $R$. It is shortly summarized as:

**Given curent tree geometry :**

$R, L$ **at each vessel** $[k, m]$

**Compute** $p^s, q^s$ **from the hemodynamic problem in a tree :**

**Get mid vessel pressure** $\bar{p}^s[k, m] = \frac{1}{2}\left\{p_T^s[k-1, \bar{m}] + p_T^s[k, m]\right\}$, $\bar{m} = \text{round}(m/2)$ (50)

**Find radius** $R[k, m]$ **minimizing cost function :**

$C(R; \bar{p}^s, q^s)[k, m] \to 0$

**Update tree geometry**